\documentclass[preprint, 1p]{elsarticle}

\usepackage{amssymb}
\usepackage{tabularray}
\usepackage{longtable}
\usepackage{xcolor}
\usepackage{pifont}
\newcommand{\cmark}{\ding{51}}
\newcommand{\xmark}{\ding{55}}
\usepackage{color}

\usepackage{background}
\usepackage{xcolor}
\usepackage{hyperref}

\journal{Computer Networks}

\hypersetup{
  colorlinks=true,
  pdfborder={0 0 0}, 
}

\backgroundsetup{
  scale=1,
  color=black,
  opacity=1,
  angle=0,
  position=current page.south,
  vshift=10pt,
  contents={\textcolor{red}{© 2024. This manuscript version is made available under the \href{https://creativecommons.org/licenses/by-nc-nd/4.0/}{CC-BY-NC-ND 4.0 license}. doi:  \href{https://doi.org/10.1016/j.comnet.2024.110215}{10.1016/j.comnet.2024.110215}}}
}

\begin{document}

\begin{frontmatter}

\title{Review and Analysis of Recent Advances in Intelligent Network Softwarization for the Internet of Things}

\author[inst1,inst2]{Mohamed Ali Zormati}

\affiliation[inst1]{organization={Heudiasyc UMR 7253, Sorbonne University Alliance, Université de Technologie de Compiègne (UTC)},
            city={Compiègne},
            country={France}}

\affiliation[inst2]{organization={National Institute of Applied Sciences and Technology (INSAT), University of Carthage},
            city={Tunis},
            country={Tunisia}}

\author[inst1]{Hicham Lakhlef}

\author[inst2]{Sofiane Ouni}

\begin{abstract}
The Internet of Things (IoT) is an emerging technology that aims to connect heterogeneous and constrained objects to each other and to the Internet. It has grown significantly in a wide variety of applications such as smart homes, smart cities, smart vehicles, etc. The huge number of connected devices increases the challenges, as IoT provides diverse and complex network services with different requirements on a common infrastructure. Network Softwarization is the latest network paradigm that transforms traditional network processes to the separation of hardware and software by using some enabling network technologies such as Software Defined Networking (SDN) and Network Function Virtualization (NFV). Machine Learning (ML) plays an essential role in creating smarter IoT networks, as it has shown remarkable results in various domains. Given that the network softwarization allows it to be easily integrated, ML can play a crucial role in efficient and self-adaptive IoT networks. \textcolor{blue}{In this paper, we provide a detailed overview of the concepts of IoT, network softwarization, and ML, and we study and discuss the state of the art of intelligent ML-enabled network softwarization for IoT.  We also identify the most prominent future research directions to be considered.}
\end{abstract}


\end{frontmatter}

\section{Introduction}
The Internet of Things (IoT) is a promising technology that creates connectivity of anything from anywhere at any time \citep{xie22}. It impacts all areas of human life as it is applied in diverse fields such as smart cities, healthcare, smart vehicles, etc. \citep{yazdinejad20} \citep{shamsan18}. It is growing exponentially \citep{ros22} as the number of IoT devices is expected to exceed 30 billion by 2025. The huge number of interconnected heterogeneous devices poses new challenges to IoT networks as these devices generate a massive and increasing amount of data \citep{shamsan18} \citep{hajian22}. The heterogeneity of IoT is also confirmed by the simultaneous use of different communication technologies (e.g., cellular networks, Wi-Fi, Zigbee), all in a large geographically distributed network. Managing such a network poses technical challenges, and to overcome these problems, IoT networks should be combined with other technologies, such as network softwarization \citep{shamsan22} \citep{shamsan19}.

Network softwarization is a novel key enabling technology to meet IoT requirements for flexibility and to build dynamic and agile IoT networks \citep{xie22}. It combines both Software Defined Networking (SDN) and Network Function Virtualization (NFV) and aims to transform the communication process and system components from legacy network devices, where software is tightly bundled with hardware \citep{amin21}, to general purpose devices. Thus, it enables a wide range of services through virtualization and programmable network, while ensuring remarkably low CAPital and OPerational EXpenditures (i.e., CAPEX and OPEX) \citep{shamsan22}.

SDN is a revolutionary networking technology that decouples the data plane (i.e., forwarding devices) from the control plane (i.e., network control logic) to provide a centralized global view of the network via a centralized SDN controller \citep{keshari23}. By centralizing device control functions, SDN enables programmable networks that are easier to manage, configure, and control \citep{xie22} \citep{shamsan22}. It is noted that SDN is the ideal solution for IoT scalability \citep{shamsan18}. As a key enabler for network softwarization, SDN has rapidly grown together with the NFV concept \citep{amin21}.

NFV decouples various network functions (e.g., firewall, load balancer) from physical network devices (i.e., proprietary hardware) and executes them using virtualization technologies \citep{xie22} \citep{saha20}. NFV virtualizes the infrastructure to deliver network services and functions using general-purpose devices, simplifying resource management, and provisioning of network functions, and scaling the capacity of a function on demand \citep{shamsan22}. Network functions are abstracted as logical entities and could be deployed as Virtualized Network Functions (VNFs) \citep{saha20}.

In other words, SDN supports easier management and setup of the network, and NFV supports easy deployment and scalability of the network \citep{ku17}. Although they do not require each other to be implemented, they are complementary and mutually reinforcing \citep{yi18}. However, to achieve SDN and NFV enabled networks, many challenges need to be fully addressed \citep{xie22}. Some potential challenges \citep{gebremariam19} are security concerns, resource utilization, fault management, etc. Machine Learning (ML) has the potential to be used to address the above issues, as ML is regarded as a useful tool to make networks self-aware, self-adaptive, self-secure, and self-managed by embedding intelligence \citep{amin21} \citep{gebremariam19}.

Until a few years ago, most networks followed a traditional approach \citep{amin21}. Recently, ML techniques have become a promising approach to bring intelligence to the network \citep{sellami20}. ML is a category of Artificial Intelligence (AI) based on intelligence that can learn from data, make decisions, identify patterns, and perform various actions with less human intervention \citep{amin21}. ML techniques allow networks to learn from experience to make them more robust against vulnerabilities and failures and to improve performance \citep{gebremariam19}. It is noted that the architectural logic of network softwarization, and particularly SDN \citep{amin21}, is better suited to ML algorithms than traditional algorithms.

\textcolor{blue}{\subsection{Contributions}}

\textcolor{blue}{To the best of our knowledge, the literature lacks a comprehensive survey on ML-enabled network softwarization for IoT. This paper aims to provide an overview of the combination of these technologies, and therefore, the main contributions can be framed as follows. Firstly, we provide an overview of IoT, network softwarization, and ML techniques as this survey encompasses a wide range of technologies. We examine the state of the art of ML-based network softwarization for constrained IoT networks. Following a comparison of existing works on the topic, we discuss the research challenges and identify the lessons learned. We finally present the most prominent future research directions.}

\textcolor{blue}{\subsection{Survey organization}}

Table \ref{tb:acr} lists the main acronyms and abbreviations used in this survey. The remainder of the paper is organized as follows. In Section 2, we present and compare the related work, \textcolor{blue}{thereby introducing the novelty of our paper}. In Section 3, we introduce the background of IoT, network softwarization, and ML techniques. In Section 4, we review how ML algorithms are applied in network softwarization and how SDN and NFV techniques are applied in IoT networks, with a focus on intelligence. \textcolor{blue}{Challenges and lessons learned are discussed in Section 5, followed by the identification of key future research directions which are presented in Section 6. We conclude this paper in Section 7.}

\textcolor{blue}{\section{Related Surveys}}
 \textcolor{blue}{In recent years, IoT, ML techniques, and network softwarization (mainly SDN and NFV techniques) have received a lot of attention from academia. In this section, we review and discuss the recent related surveys in this area.}

In \citep{bizanis16}, the authors review the state of the art of applying SDN and network virtualization to IoT. They note that prior to their work, there has been no effort to survey the combination of these three technologies. The authors also present a summary of the most prominent architectural frameworks. They identify the following research challenges and future directions: optimizing SDN and \textit{OpenFlow} to adapt to the peculiarities of the IoT paradigm, considering network slicing, considering security challenges, etc. The authors conclude by suggesting to consider the merging of networking and AI.

In \citep{shamsan18}, the authors review SDN-enabled IoT architectures proposed by different researchers. They classify the architectures into three categories: layer-based (device layer, communication layer, computing layer (SDN controller), and service layer), agent-based (each device has an IoT agent to interact with IoT controllers), and domain-based (multiple SDN domains with one or more controllers in each domain). The authors conclude their survey by pointing out that more research should be published to integrate IoT with different technologies to make it more flexible and scalable, and to simplify management and control. We note that no simulation or evaluation has been done to compare the different architectural approaches. It is recommended to know which integration is better given the constraints of the IoT networks (e.g., energy limits). The authors did not consider NFV technology at this point.

The authors of the previously cited survey, reviewed in \citep{shamsan19} the network softwarization for IoT with the adoption of IoT softwarization architecture, based on both SDN and NFV. Here, SDN centrally orchestrates IoT network flows, while NFV supports the provision of IoT network services on demand. They propose a four-layer architecture: physical infrastructure layer (IoT and VNF domains), control layer (IoT and SDN controllers), orchestration layer (contains the management layer components of a virtualization system), and application layer (contains the applications useful to end users). The authors discuss their proposal, but do not present its performance evaluation, which prevents comparison with a traditional IoT architecture.

\begin{longtblr}[
  label = tb:acr,
  caption = {List of Acronyms and Abbreviations},
]{
  width = \linewidth,
  colspec = {Q[108]Q[362]Q[108]Q[362]},
  hlines,
}
\textbf{Label} & \textbf{Description} & \textbf{Label} & \textbf{Description}\\
AI & Artificial Intelligence & MDP & Markov Decision Process\\
ANN & Artificial Neural Network & MDS & Markov Decision Support\\
API & Application Programming Interface & MitM & Man in the Middle\\
BMU & Best Matching Unit & ML & Machine Learning\\
CAPEX & CAPital EXpenditures & MLP & MultiLayer Perceptron\\
CFL & Centralized Federated Learning & NFV & Network Function Virtualization\\
CNN & Convolutional Neural Network & NFVI & Network Function Virtualization Infrastructure\\
DAG & Directed Acyclic Graph & OPEX & Operational Expenditures\\
DDoS & Distributed Denial of Service & P2P & Peer to Peer\\
DFL & Decentralized Federated Learning & PDR & Packet Delivery Ratio\\
DL & Deep Learning & QoE & Quality of Experience\\
DNN & Deep Neural Network & QoS & Quality of Service\\
DRL & Deep Reinforcement Learning & RF & Random Forest\\
EC & Edge Computing & RL & Reinforcement Learning\\
FDRL & Federated Deep Reinforcement Learning & RNN & Recurrent Neural Network\\
FL & Federated Learning & RPL & Routing Protocol for Low-power and lossy networks\\
FRL & Federated Reinforcement Learning & SDN & Software Defined Networking\\
FTL & Federated Transfer Learning & SFC & Service Funtion Chain\\
GMM & Gaussian Mixture Model & SL & Supervised Learning\\
GNN & Graph Neural Network & SOM & Self-Organizing Map\\
HFL & Horizontal Federated Learning & SPoF & Single Point of Failure\\
IoT & Internet of Things & UL & Unsupervised Learning\\
\textit{k}-NN & \textit{k}-Nearest Neighbor & VFL & Vertical Federated Learning\\
MANO & Management ANd Operation & VNF & Virtualized Network Function\\
\end{longtblr}

\newpage

In \citep{gebremariam19}, the authors review the main application areas of AI and ML techniques in SDN- and NFV-based networks. They categorize their applications into specific tracks, including network architecture, network planning, network management and operation, and network security. They identify and discuss the challenges and most prominent future directions in the field. The authors point out that the key challenges are: computational complexity and latency, computational and storage resource requirements, access to resources and datasets (there is a lack of openly available standard network datasets), and storage of valuable data (there is no trend to store communication data for future use). Conversely, we note that the authors did not execute the studied methods, but plan as future work to quantify the resource usage of network elements by executing the ML algorithms for different network scales and traffic loads.

In \citep{kellerer19}, the authors review and examine the opportunities and challenges of adaptive and data-driven softwarized networks, and introduce a conceptual framework for adaptation (i.e., flexible response to new requirements and changing contexts due to the diversity of applications) in softwarized networks. The data-driven decision modules (i.e., ML modules) can learn and make informed decisions in response to environmental changes, thereby facilitating meaningful decision-making for adaptation in softwarized networks. We note that the authors do not address the IoT scenario.

In \citep{xie19}, the authors provide a comprehensive review of the literature on ML applied to SDN from the perspective of traffic classification (perform fine-grained network management by identifying different traffic flow types), routing optimization (where the use of Reinforcement Learning (RL) outperforms the conventional heuristic algorithms), Quality of Service (QoS) and Quality of Experience (QoE) prediction, resource management, and security. The authors also discuss the challenges, mainly mentioning the need for high-quality standardized datasets. In summary, the research on the application of ML algorithms is quite broad and many challenges lie ahead. Even if the authors compare the performance of ML-based solutions in SDN in terms of accuracy, they could not evaluate them directly, as different learning algorithms use different datasets for training and simulation is not performed under the same conditions.

While reference \citep{xie19} provides a general survey, the authors in \citep{amin21} provide a specific survey, focusing on the application of ML for optimizing routing in SDN-enabled environments. They highlight the ongoing need for more extensive comparisons and collaborations among different approaches, as well as the necessity for meaningful evaluations based on openly available datasets and network topologies. Additionally, the authors acknowledge the challenge of acquiring large datasets and suggest addressing it by initially training models with synthetic data followed by fine-tuning using smaller real datasets. The authors conclude that Deep Reinforcement Learning (DRL) is particularly relevant in the last two years, as most of the published works fall into this type of ML technique. However, there is a lack of efforts to create synergies or to compare different ML approaches for SDN routing. Many evaluations only compare their methods with traditional routing protocols and not with competing proposals, likely due to limited public availability of implementations. This lack of reproducibility and comparability hinders the meaningfulness and conclusiveness of these evaluations. We also note that the authors excluded Federated Learning (FL) from the classification, even though it is appropriate for networking, mainly because the authors did not find any work using FL for routing optimization in SDN.

~

In \citep{alonso21}, one of the rare surveys that focuses on the combination of ML, SDN, NFV and IoT (although it does not cover all ML approaches, contrary to ours), the authors present an overview of using DRL for managing SDN and NFV in Edge-IoT scenarios. It is worth highlighting that Edge Computing (EC) architecture plays a pivotal role in delivering quicker service response times and cost reduction in processing IoT data compared to traditional cloud-based approaches. As the authors noted, the works do not focus on RL solutions, or more specifically, DRL, or do not go into depth, and in some cases, they deal with them only superficially. Unfortunately, we note that no simulation or evaluation has been performed to compare the existing solutions.

The authors in \citep{alam21} propose a systematic and comprehensive survey of virtualization techniques explicitly designed for IoT networks. The solutions are detailed in three categories: architectural, security, and management solutions. The authors highlight short- and long-term research challenges. We note that there is no special focus on intelligent solutions.

In \citep{abid22}, the authors discuss the traditional IoT networks and the need for SDN and NFV to address IoT challenges. They acknowledge that a lot of research has been done in the areas of IoT, SDN, and NFV, but they note that there is a missing link in terms of the evolution of IoT architectures, from primitive frameworks to sophisticated SDN- and NFV-enabled platforms. The proposed survey attempts to fill this gap.  The authors superficially discuss the application of AI and ML techniques in IoT (i.e., cognitive IoT networks) and in IoT network softwarization. The authors explore the architectural evolution of IoT. They also survey the state of the art of the available IoT simulators. We note that ML techniques have not been given much attention, as they are required as promising technologies to achieve efficient IoT network softwarization in the near future.

\textcolor{blue}{The authors in \citep{javanmardi23} review the use of SDN in IoT fog-enabled networks, with the specific goal of improving security and countering cyber threats. They recall that although the use of fog computing in IoT has optimized resource utilization in such a constrained environment, it has raised more concerns about IoT security (e.g., at the fog layer). Since achieving security in IoT fog networks requires a broad perspective, it is interesting to consider SDN. After identifying the advantages of SDN in the studied context, analyzing the vulnerabilities arising from the combination of IoT and fog computing, and exploring the SDN-based security measures, the authors review the most recent and relevant security mechanisms for IoT fog networks. Compared to the existing surveys, the novelty introduced in this work is to examine how SDN improves fog security in IoT networks. However, NFV is not considered, even though it is highly recommended to improve security in softwarized networks. Although the authors mention the importance of using intelligent learning approaches, this is not considered as a comparison criterion in this review article.}

\textcolor{blue}{In \citep{turner23}, the authors propose a comprehensive review of the integration of SDN and blockchain into the IoT ecosystem to improve security and network performance. The scope of this study revolves around the integration of IoT, SDN, and blockchain. However, it does not consider NFV as a softwarization enabler, although it is a major key technology for achieving security in combination with SDN. Although the integration of ML is not explored in this study, the authors suggest the combination of IoT, SDN, and blockchain with ML (especially FL) as a future research direction that needs further investigation.}

\begin{longtblr}[
  label = tb:surv,
  caption = {A Brief Comparison of Our Survey with Existing Ones},
]{
  width = \linewidth,
  colspec = {Q[63]Q[131]Q[119]Q[131]Q[125]Q[48]Q[50]Q[50]Q[46]Q[150]},
  row{2} = {c},
  column{10} = {c},
  cell{1}{1} = {r=2}{},
  cell{1}{2} = {r=2}{c},
  cell{1}{3} = {r=2}{c},
  cell{1}{4} = {c=2}{0.256\linewidth,c},
  cell{1}{6} = {c=4}{0.194\linewidth,c},
  cell{1}{10} = {r=2}{},
  cell{3}{4} = {c},
  cell{3}{5} = {c},
  cell{3}{6} = {c},
  cell{3}{7} = {c},
  cell{3}{8} = {c},
  cell{3}{9} = {c},
  cell{4}{4} = {c},
  cell{4}{5} = {c},
  cell{4}{6} = {c},
  cell{4}{7} = {c},
  cell{4}{8} = {c},
  cell{4}{9} = {c},
  cell{5}{4} = {c},
  cell{5}{5} = {c},
  cell{5}{6} = {c},
  cell{5}{7} = {c},
  cell{5}{8} = {c},
  cell{5}{9} = {c},
  cell{6}{4} = {c},
  cell{6}{5} = {c},
  cell{6}{6} = {c},
  cell{6}{7} = {c},
  cell{6}{8} = {c},
  cell{6}{9} = {c},
  cell{7}{4} = {c},
  cell{7}{5} = {c},
  cell{7}{6} = {c},
  cell{7}{7} = {c},
  cell{7}{8} = {c},
  cell{7}{9} = {c},
  cell{8}{4} = {c},
  cell{8}{5} = {c},
  cell{8}{6} = {c},
  cell{8}{7} = {c},
  cell{8}{8} = {c},
  cell{8}{9} = {c},
  cell{9}{4} = {c},
  cell{9}{5} = {c},
  cell{9}{6} = {c},
  cell{9}{7} = {c},
  cell{9}{8} = {c},
  cell{9}{9} = {c},
  cell{10}{4} = {c},
  cell{10}{5} = {c},
  cell{10}{6} = {c},
  cell{10}{7} = {c},
  cell{10}{8} = {c},
  cell{10}{9} = {c},
  cell{11}{4} = {c},
  cell{11}{5} = {c},
  cell{11}{6} = {c},
  cell{11}{7} = {c},
  cell{11}{8} = {c},
  cell{11}{9} = {c},
  cell{12}{4} = {c},
  cell{12}{5} = {c},
  cell{12}{6} = {c},
  cell{12}{7} = {c},
  cell{12}{8} = {c},
  cell{12}{9} = {c},
  cell{13}{1} = {fg=blue},
  cell{13}{2} = {fg=blue},
  cell{13}{4} = {c},
  cell{13}{5} = {c},
  cell{13}{6} = {c},
  cell{13}{7} = {c},
  cell{13}{8} = {c},
  cell{13}{9} = {c},
  cell{14}{1} = {fg=blue},
  cell{14}{2} = {fg=blue},
  cell{14}{4} = {c},
  cell{14}{5} = {c},
  cell{14}{6} = {c},
  cell{14}{7} = {c},
  cell{14}{8} = {c},
  cell{14}{9} = {c},
  cell{15}{4} = {c},
  cell{15}{5} = {c},
  cell{15}{6} = {c},
  cell{15}{7} = {c},
  cell{15}{8} = {c},
  cell{15}{9} = {c},
  hline{1,4-16} = {-}{},
  hline{2} = {4-9}{},
  hline{3} = {4-10}{},
}
\textbf{Year} & \textbf{Paper Type} & \textbf{Ref.} & \textbf{Network Softwarization} &  & \textbf{ML Technique(s)} &  &  &  & \textbf{IoT Networks}\\
 &  &  & \textbf{SDN} & \textbf{NFV} & \textbf{SL} & \textbf{UL} & \textbf{RL} & \textbf{FL} & \\
2016 & Journal & \cite{bizanis16} & \cmark & \xmark & \xmark & \xmark & \xmark & \xmark & \cmark\\
2018 & Conference & \cite{shamsan18} & \cmark & \xmark & \xmark & \xmark & \xmark & \xmark & \cmark\\
2019 & Journal & \cite{shamsan19} & \cmark & \cmark & \xmark & \xmark & \xmark & \xmark & \cmark\\
2019 & Conference & \cite{gebremariam19} & \cmark & \cmark & \cmark & \cmark & \cmark & \xmark & \xmark\\
2019 & Journal & \cite{kellerer19} & \cmark & \cmark & \xmark & \xmark & \xmark & \xmark & \xmark\\
2019 & Journal & \cite{xie19} & \cmark & \xmark & \cmark & \cmark & \cmark & \xmark & \xmark\\
2021 & Journal & \cite{amin21} & \cmark & \xmark & \cmark & \cmark & \cmark & \xmark & \xmark\\
2021 & Workshop & \cite{alonso21} & \cmark & \cmark & \xmark & \xmark & \cmark & \xmark & \cmark\\
2021 & Journal & \cite{alam21} & \cmark & \cmark & \xmark & \xmark & \xmark & \xmark & \cmark\\
2022 & Journal & \cite{abid22} & \cmark & \cmark & \xmark & \xmark & \xmark & \xmark & \cmark\\
\textcolor{blue}{2023} & \textcolor{blue}{Journal} & \textcolor{blue}{\cite{javanmardi23}} & \textcolor{blue}{\cmark} & \textcolor{blue}{\xmark} & \textcolor{blue}{\xmark} & \textcolor{blue}{\xmark} & \textcolor{blue}{\xmark} & \textcolor{blue}{\xmark} & \textcolor{blue}{\cmark}\\
\textcolor{blue}{2023} & \textcolor{blue}{Journal} & \textcolor{blue}{\cite{turner23}} & \textcolor{blue}{\cmark} & \textcolor{blue}{\xmark} & \textcolor{blue}{\xmark} & \textcolor{blue}{\xmark} & \textcolor{blue}{\xmark} & \textcolor{blue}{\xmark} & \textcolor{blue}{\cmark}\\
\textbf{2023} & \textbf{Journal} & \textbf{Ours} & \cmark & \cmark & \cmark & \cmark & \cmark & \cmark & \cmark

\end{longtblr}

As summarized in Table \ref{tb:surv} (where we compare our work with the existing survey papers presented above), we find that although many research efforts have been made to review the three promising technologies (IoT, ML, and network softwarization) and their combination, only a few of them have considered combining all these three technologies, but within a narrow vision (e.g., considering only one ML technique). Moreover, to the best of our knowledge, no existing work has really focused on ML-enabled network softwarization for IoT networks as a promising solution to address multiple challenges. It is also to note that there is no work that consider FL technique as an enabler for efficient network softwarization, especially for IoT networks. 

To fill this gap, in this paper, we provide a comprehensive overview of the intelligent network softwarization for IoT constrained networks. We hope that our discussion and exploration can provide an overall understanding of this emerging field, and encourage more subsequent studies on the topic.

\section{Background}
In this section, we present the main concepts of IoT, network softwarization with an emphasis on SDN and NFV technologies, and ML techniques. \textcolor{blue}{We highlight some interesting aspects of each technology, and we identify the potential advantages and challenges of integrating SDN, NFV, and ML with IoT.}

\subsection{Internet of Things (IoT)}

The Internet of Things (IoT) connects billions of devices (i.e., things) to provide a wide array of services by establishing connectivity between these devices and the Internet. \textcolor{blue}{In addition to connecting things, IoT also focuses on optimizing traditional systems, as evidenced by the various IoT applications \citep{sobin20} \citep{abdalzaher23}. As examples of current and cutting edge IoT use cases, we cite the use of IoT in environmental monitoring, where it enables the detection and prevention of natural disasters, and its use in agriculture, where it helps automate manual processes (e.g., irrigation) while giving farmers visibility to optimize resources.}

\begin{figure}[htbp]
\centerline{\includegraphics[scale=0.46]{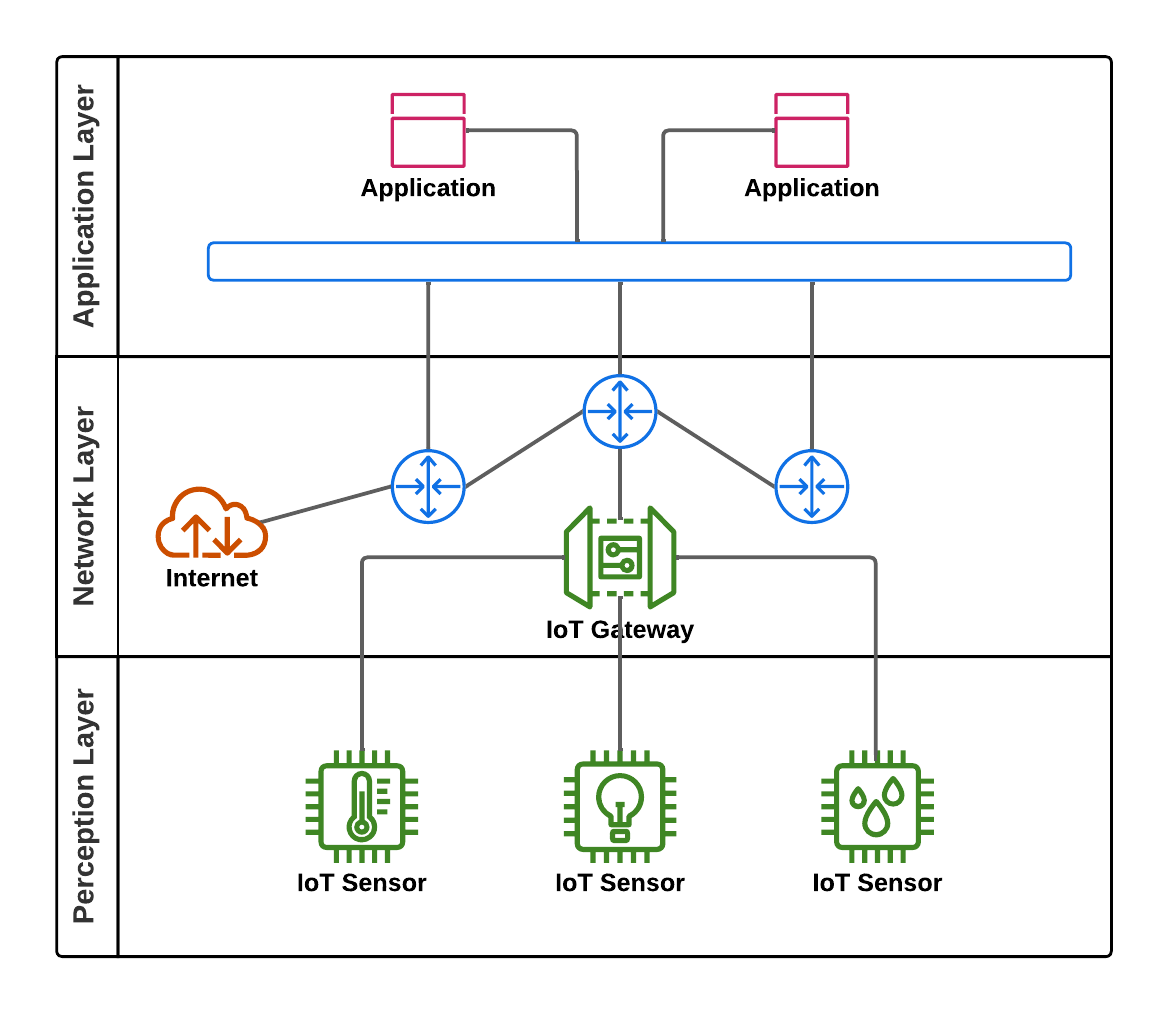}}
\caption{IoT Architecture}
\label{iotarch}
\end{figure}

Several architectures have been proposed for IoT \citep{al_qaseemi16}. According to \citep{shamsan19}, most of the proposals consist of three layers, as shown in Figure \ref{iotarch}.

\begin{itemize}
    \item \textbf{Perception layer:} Contains the objects used for sensing. This layer is responsible for sensing the environment, collecting the sensing data, and sending it to the next layer (specifically, to the IoT gateway).
    \item\textbf{Network layer:} Contains the network devices and communication technologies. This layer is considered the interface between perception and application layers.
    \item\textbf{Application layer:} Contains the IoT services and applications. This is the front end of the system as it is an interface between the IoT system and the users.
\end{itemize}

We note that some architectures add a fourth layer, called the \textbf{Middleware layer}. It coordinates between IoT devices provided by different vendors, thus enabling communication between all objects regardless of their hardware \citep{shamsan19}. This provides more flexibility in the association between devices \citep{al_qaseemi16}. The middleware layer is in charge of service management over IoT devices to establish connections between the ones that provide the same service \citep{mousavi21}.

The complexity of IoT is mainly due to the fact that a large number of things are connected to the Internet and communicate with each other, while being governed by different protocols and models \citep{shamsan19}. In fact, IoT faces several challenges and has specific requirements, which are detailed below \citep{sobin20} \citep{al_qaseemi16} \citep{mousavi21} \citep{farhan17}:
\begin{itemize}
    \item \textbf{Heterogeneity:} It comes into picture because IoT can connect any physical object. Thus, a unifying architecture is needed to implement IoT correctly.
    \item \textbf{Scalability:} It needs to be considered while designing solutions for IoT (e.g., routing protocols and data storage mechanisms).
    \item \textbf{Software development challenges:} Consider the 5Vs of big data (volume, variety, velocity, veracity, and value) and self-configuration (software, hardware, and network configuration).
    \item \textbf{Standardization:} Currently, there is a lack of standardization, which limits the effectiveness and performance of the IoT.
    \item \textbf{Security:} For example, IoT systems need to consider the necessary criteria for establishing computer system security (e.g., confidentiality, integrity, authentication, non-repudiation, availability, etc.).
\end{itemize}

To address the above issues and challenges, and to meet the requirements of IoT, many research works propose network softwarization as a key enabling technology for IoT networks.

\subsection{Network softwarization}
\textcolor{blue}{Network softwarization is an emerging approach \cite{yi18} to transform the traditional networks to the new trending technologies such as programmable network and virtualization. By leveraging the software features such as flexibility, it aims to provide new functionalities and deploy them efficiently \cite{shamsan19}.} 

\textcolor{blue}{The goal is to transform the network into an open ecosystem that decouples hardware and software components. This transformation is expected to deliver benefits such as advanced services, enhanced networking capabilities, and improved network development and maintenance capabilities \cite{popescu22}. The following features show the importance of network softwarization: separation of responsibilities (by separating data and control planes), configurability and global view (by having a centralized controller), scalability (the ability to add and dynamically configure network functions as needed), cost efficiency (reduced CAPEX and OPEX), etc.}

Both SDN and NFV technologies play an important role in enabling network softwarization and increasing programmability \citep{shamsan19}, along with other emerging technologies (e.g., cloud computing, edge computing, network slicing) \citep{popescu22}.

\subsubsection{Software Defined Networking (SDN)}~\par
\textcolor{blue}{The last decade has seen a new wave of networking innovation, largely due to the Software Defined Networking (SDN) paradigm \cite{amin21}. SDN is currently being deployed in all cloud computing environments and server operations, and is potentially serving as the foundation for the majority of future network services \cite{shamsan18}. The key feature of SDN is the separation of the network control plane (responsible for routing) from the data plane (responsible for forwarding) \cite{ouhab20}. SDN provides a set of Application Programming Interfaces (APIs) that can implement network services for business purposes \cite{shuker19}.}

\textcolor{blue}{By decoupling the control plane and data plane, SDN improves the network architecture and eliminates its hierarchy \cite{shamsan19}. Network devices (e.g., routers, switches, access points) become forwarding devices. The centralized controller is responsible for managing and controlling all network functions and can dynamically program the network \cite{xie19} \cite{mijumbi16}. SDN's centralized architecture provides a faster view of network status, enables more straightforward programmability and updates \cite{amin21}, and has the potential to increase network flexibility and performance \cite{keshari23}. SDN is a key solution to ensure network configurability, global knowledge, and network virtualization \cite{bizanis16} \cite{baddeley18}.}

SDN faces several challenges that need to be addressed for effective implementation. The most important ones are listed below \citep{amin21}:

\begin{itemize}
    \item \textbf{Controller location:} For instance, SDN introduces an additional communication channel between the data plane and the control plane, which may lack complete transparency, especially in large networks.
    \item \textbf{Scalability:} To avoid bottlenecks, administrators should consider how much control should be delegated to the controller.
    \item \textbf{Security:} This challenges all networks, but SDN-enabled networks are more vulnerable because they are logically centralized.
    \item \textbf{Interoperability:} This is especially relevant in large networks. In addition to device heterogeneity, interoperability between different SDN technologies remains a challenge.
    \item \textbf{Reliability:} This challenge is exacerbated in SDN-enabled networks because the control communication represents a new potential point of failure that should be protected.
\end{itemize}

\begin{figure}[htbp]
\centerline{\includegraphics[scale=0.4]{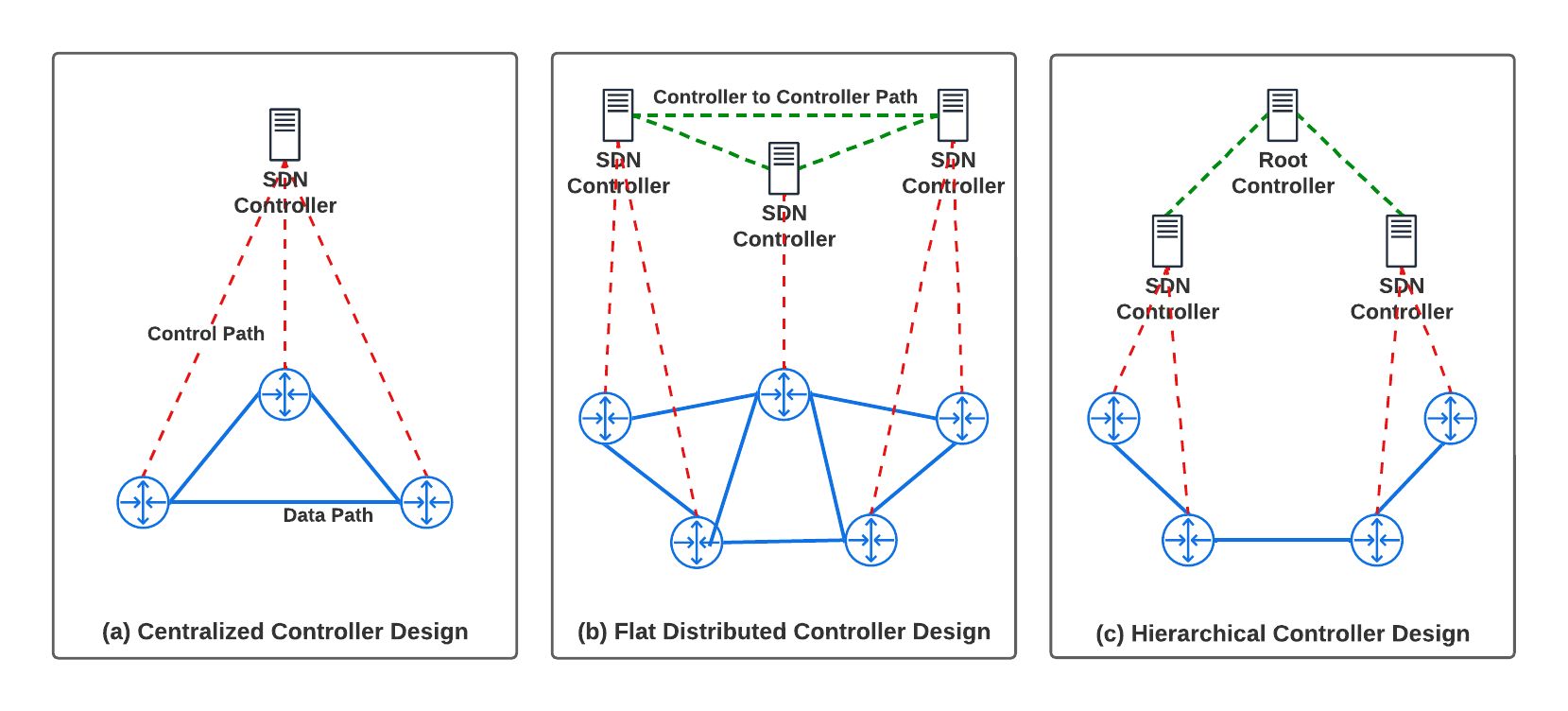}}
\caption{SDN Controller Design}
\label{contdesign}
\end{figure}

In SDN, a single controller is typically responsible for the entire system. To address many of the possible issues, SDN can distribute the control plane by having multiple controllers \citep{shamsan18}. As shown in Figure \ref{contdesign}, approaches can be centralized or distributed \citep{shirmarz20}. Although the most common case is to have one controller working in a centralized manner (a), it is possible to have a flat distributed controller design (b), or a hierarchical controller design (c).

\begin{figure}[htbp]
\centerline{\includegraphics[scale=0.45]{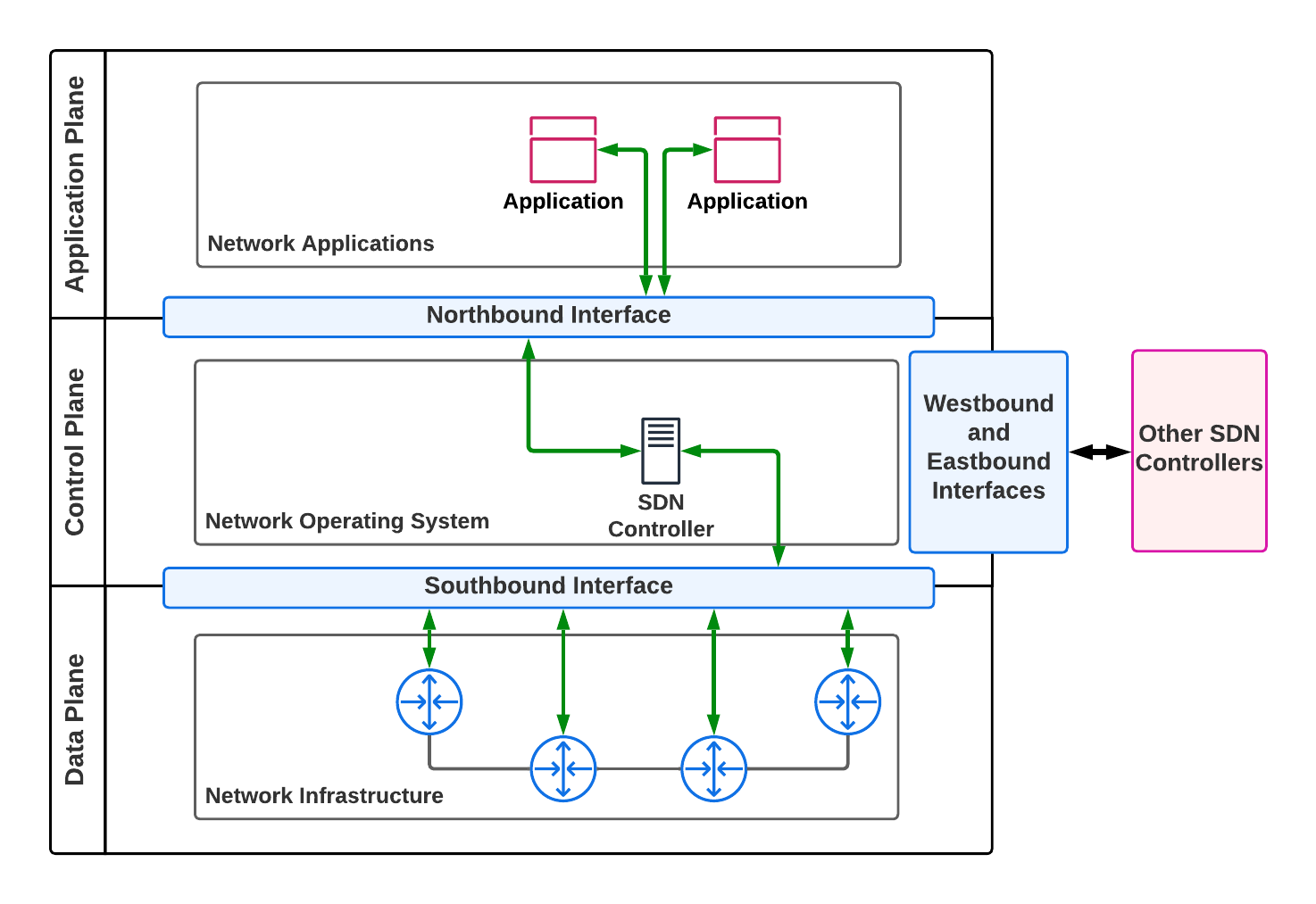}}
\caption{SDN Architecture}
\label{sdnarch}
\end{figure}

The architecture of SDN is depicted in Figure \ref{sdnarch}. It consists of three planes that can communicate through APIs \citep{shamsan19} \citep{amin21} \citep{xie19}.

\begin{itemize}
    \item \textbf{Data plane:} It consists of the set of physical or virtual network devices. It handles incoming frames (i.e., forwarding, modifying, or discarding the frame) according to the policies of the control plane.
    \item \textbf{Control plane:} This is the \textit{brain} of the network, responsible for making decisions and making network management agile. It consists of core network functions that are common to all types of applications (e.g., network topology discovery). Its main component is the logically centralized SDN controller, which provides network abstraction and a global view of the topology and its components.
    \item \textbf{Application plane:} It encompasses business applications designed to meet user requirements and plays a crucial role in defining the network's overall behavior according to the preferences of the network administrator.
\end{itemize}

There are four types of interfaces in SDN that are used to facilitate communication between different layers and multiple components \citep{shamsan19}. Southbound interface is used to ensure communication between data plane and control plane. It should be noted that \textit{OpenFlow} is considered as a best practice and example of southbound APIs, as it is used to get state and statistics from forwarding elements and give them control directives \citep{shuker19}. Northbound interface is used as an interface between the control plane and the application plane. It provides the view of the network to the applications. To ensure redundancy, there should be multiple SDN controllers that coordinate decisions and communicate with each other through westbound and eastbound interfaces \citep{shamsan18}.

\subsubsection{Network Function Virtualization (NFV)}~\par
SDN has rapidly grown together with the Network Function Virtualization (NFV) concept, as they combined forces to boost emergent networking applications \citep{amin21}. NFV is a key enabling technology to deliver on-demand network services \citep{shamsan22}.

NFV transfers the network functions from specific hardware to software virtualized platforms. In that, the network devices are hosted in general-purpose hardware using virtualization technologies \citep{shamsan19}. Thus, network functions become Virtual Network Functions (VNFs) that are implemented on multiple components over multiple virtual machines.

It should be emphasized that NFV has several significant benefits \cite{shamsan19}. It enables increased network and resource flexibility, reduced time to introduce new services, improved resource utilization (which implies cost efficiency), and ensuring QoS and QoE. To achieve these benefits and thus solve most of the current network problems \cite{gilherrera16}, NFV introduces several changes in network service provisioning compared to conventional approaches. In summary, these differences are as follows \cite{mijumbi16}: decoupling of software from hardware, enabling flexible provisioning of network functions, and facilitating dynamic scaling.

\begin{figure}[htbp]
\centerline{\includegraphics[scale=0.42]{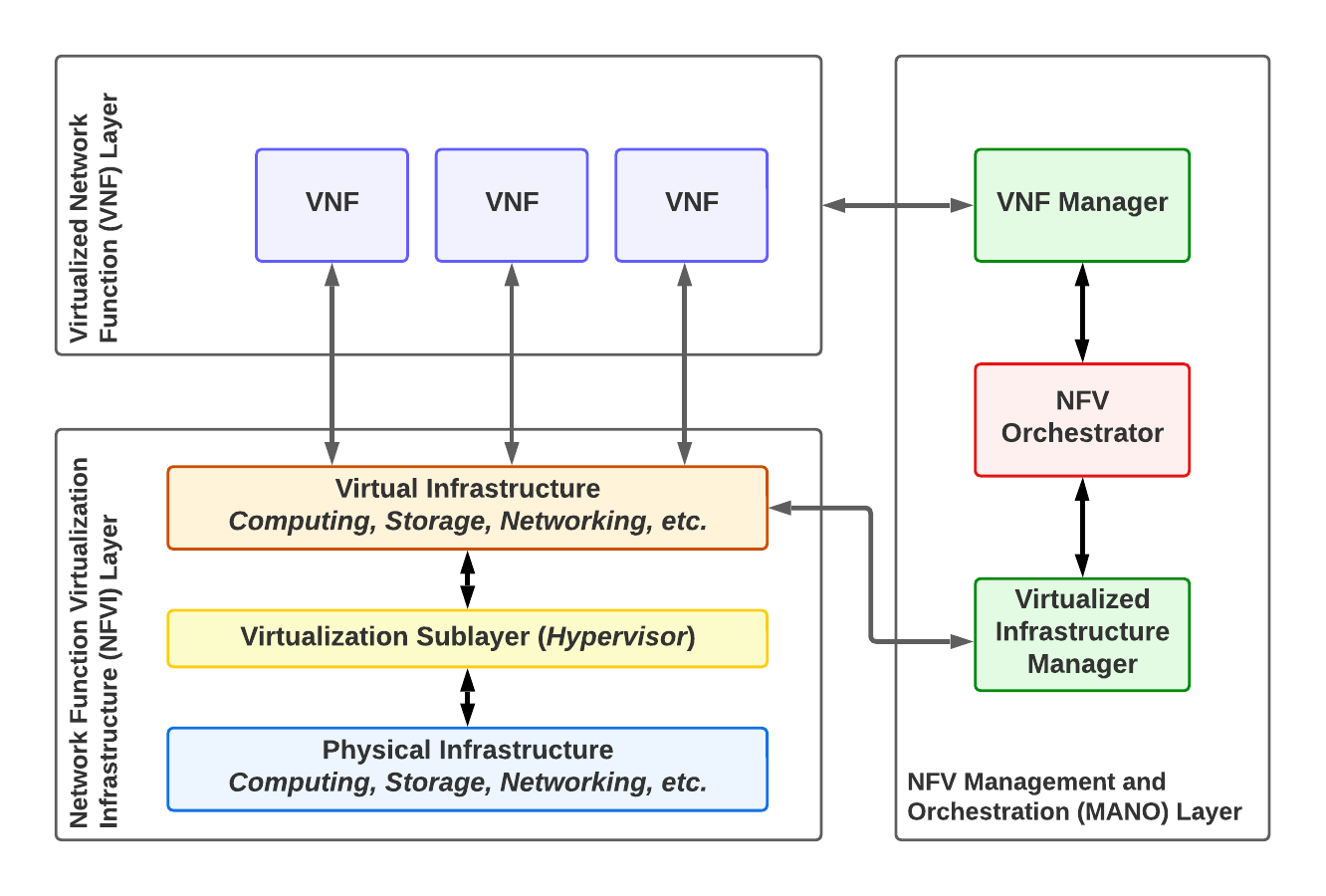}}
\caption{NFV Architecture}
\label{nfvarch}
\end{figure}

Figure \ref{nfvarch} introduces the NFV architecture, that is composed of three main layers that are described below \cite{shamsan19}.

\begin{itemize}
    \item \textbf{Network Function Virtualization Infrastructure (NFVI) layer:} It provides the infrastructure of general-purpose physical devices and virtualization environment. It contains three sublayers: physical infrastructure (i.e., hardware components), virtualization sublayer (i.e., the hypervisor), and virtual infrastructure (i.e., the virtualization environment).
    \item \textbf{NFV Management and Orchestration (MANO) layer:} It is responsible for managing all virtualization processes of NFV. It includes three major categories of management: virtualized infrastructure manager (manages the resources of NFVIs), VNF manager (manages multiple VNFs), and NFV orchestrator (manages the VNFs lifecycle and orchestrates the resources of NFVIs).
    \item \textbf{Virtual Network Function (VNF) layer:} It is responsible for providing functionalities and services on general purpose.
\end{itemize}

Admittedly, NFV is still in its early stages, as many key challenges need to be fully addressed \citep{laghrissi19}, and one of the most important challenges is how to optimally allocate virtual resources to network services \citep{xie22}. Aspects such as management and orchestration, energy efficiency, security, and modeling of resources, functions, and services should also be explored and adequately addressed. It is also important to note that the coexistence of NFV and legacy systems is inevitable before the complete transition to NFV, since it is economically and practically unfeasible to virtualize all functions immediately \cite{yi18}.

\subsection{Machine Learning (ML)}
Thanks to network softwarization, the application of Artificial Intelligence (AI) and Machine Learning (ML) to networking is easier to implement nowadays \citep{amin21}. ML is a branch of AI that defines any computational method in which the results of previous events or decisions are used to improve predictions or decisions \citep{memos22}, allowing systems to improve themselves without being explicitly programmed.

\begin{figure}[htbp]
\centerline{\includegraphics[scale=0.4]{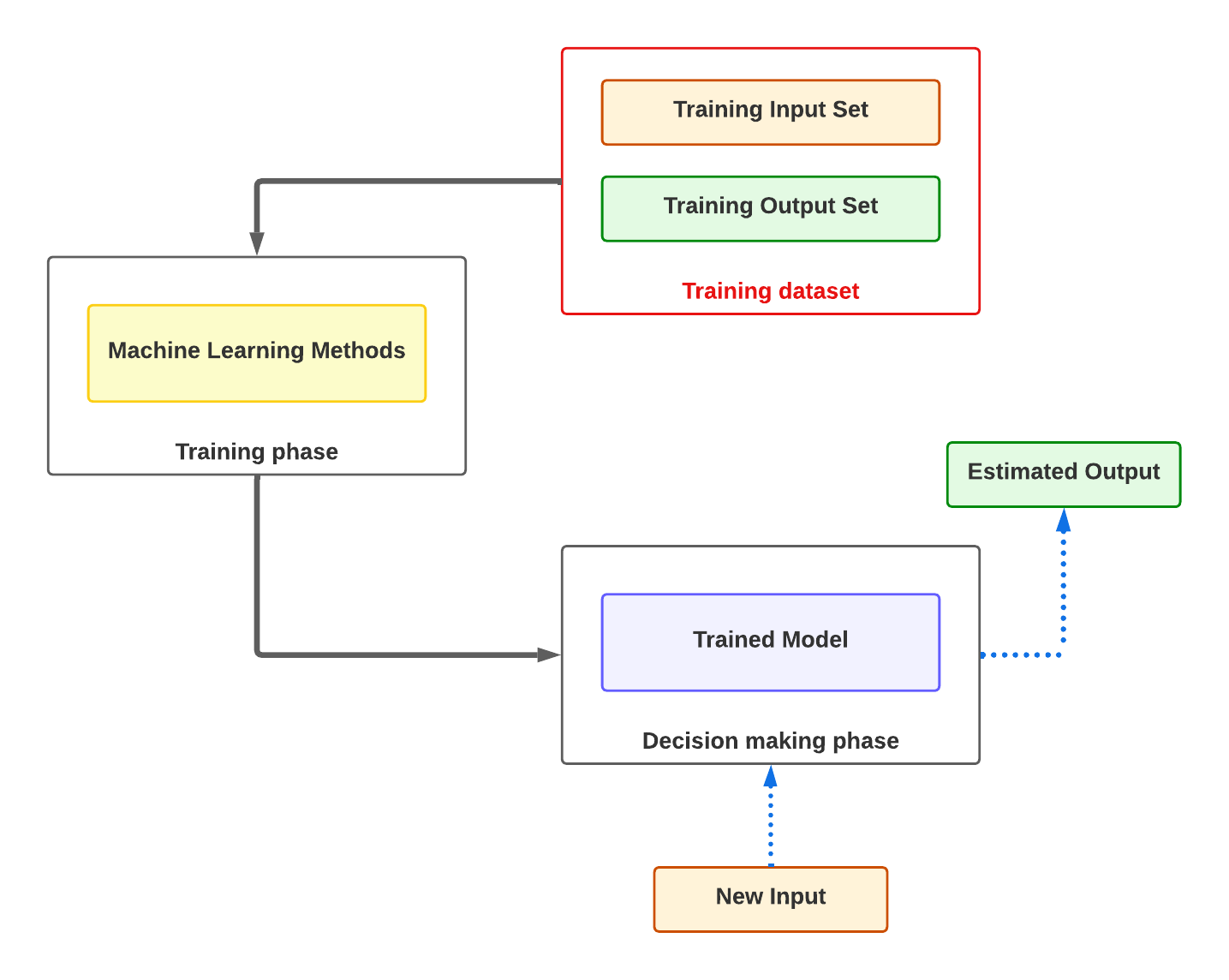}}
\caption{Main Phases of an ML Approach}
\label{phasesml}
\end{figure}

ML approaches typically consist of two main phases, as depicted in Figure \ref{phasesml}: a training phase and a decision making phase \citep{xie19}. In the training phase, ML methods are employed to acquire knowledge from a training dataset and construct a system model. In the subsequent decision-making phase, the trained model is utilized to generate estimated outputs for new inputs, facilitating informed decision-making.

\textcolor{blue}{ML is a very large field whose methods have been classified into several categories. We propose a classification approach based on the type of learning involved, thus distinguishing the supervised, unsupervised, and reinforcement learning, and we introduce below many widely used ML algorithms \cite{amin21} \cite{xie19}. We also introduce federated learning as an emerging orthogonal learning paradigm \cite{amin21}.}

\subsection{Supervised Learning (SL)}~\par
Supervised Learning (SL) revolves around identifying the unknown function that links the input and output spaces, using labeled input-output pairs for a process known as training. Once the training is complete, the trained model can predict the expected output when provided with new input data.

We present below the most commonly used supervised methods, although certain techniques can also be adapted for unsupervised learning or incorporated into a reinforcement learning framework \cite{amin21}.

\begin{itemize}
    \item \textcolor{blue}{\textbf{\textit{k}-Nearest Neighbor (\textit{k}-NN):} Classifies a data sample based on the \textit{k}-nearest neighbors of that unclassified sample.}
    \item \textcolor{blue}{\textbf{Linear Regression:} Is a highly effective ML approach that assumes a linear relationship between the dependent variable and the independent variables.}
    \item \textcolor{blue}{\textbf{Markov Decision Process (MDP):} Is a stochastic process operating in discrete time, characterized by the Markov property, which means that the probability of transitioning to a specific state depends solely on the current state.}
    \item \textcolor{blue}{\textbf{Decision Tree:} Employs a learning tree structure where each node represents a feature of the data, branches symbolize conjunctions of features leading to classifications, and leaf nodes correspond to class labels. Unlabeled samples are classified by comparing their feature values with the nodes of the decision tree.}
    \item \textcolor{blue}{\textbf{Random Forests (RF):} Is a method that aggregates the results of numerous decision trees to estimate a unique value in regression or to determine a class in classification. Each tree contributes a classification result (a vote), and the data sample is classified into the class with the most votes.}
    \item \textcolor{blue}{\textbf{Artificial Neural Network (ANN):} Consists of interconnected units (i.e., artificial neurons) that use activation functions to perform nonlinear computations and extract knowledge from historical data. Perceptron and MultiLayer Perceptron (MLP) were the initial architectures of ANNs.}
    \item \textcolor{blue}{\textbf{Deep Neural Network (DNN):} Also known as Deep Learning (DL), this category of ANNs encompasses network architectures with a high number of interconnected layers. Convolutional Neural Network (CNN) and Recurrent Neural Network (RNN) are two prominent types of DNNs. CNN is a feed-forward network composed of a sequence of cascading convolutional layers. RNN is a stateful network capable of utilizing internal state to handle sequential data.}
\end{itemize}

\subsection{Unsupervised Learning (UL)}~\par
Unsupervised Learning (UL) is designed to identify patterns within unlabeled datasets, eliminating the need for human supervision and pre-labeled input-output pairs. Unsupervised methods autonomously infer relationships between variables based on features such as correlations. These techniques are commonly employed in clustering and data aggregation, grouping sample data into distinct clusters based on their similarities. The most common UL methods are presented below \cite{amin21} \cite{xie19}.

\begin{itemize}
    \item \textbf{\textit{k}-Means:} Stands as a widely adopted UL algorithm employed to cluster a set of data observations into \textit{k} clusters. This technique works by minimizing the variance within these clusters. Each observation is allocated to the cluster whose centroid is closest in terms of distance.
    \item \textbf{Gaussian Mixture Models (GMM):} Assumes that observations are generated by a mixture of a finite number of Gaussian variables. This model is probabilistic in nature and extends the \textit{k}-means approach by incorporating covariance, thereby capturing the uncertainty in cluster assignments.
    \item \textbf{Hierarchical Clustering:} Assembles nearby observations into clusters and establish links. The outcome is a partially ordered dendrogram that gives a hierarchy to the clusters.
    \item \textbf{Self-Organizing Maps (SOM):} Are unsupervised ANNs specifically designed to uncover a low-dimensional discrete representation, known as a map, from the input space. These SOMs are trained using unlabeled data and employ competitive learning principles, wherein only one neuron, termed the Best Matching Unit (BMU), is selected at each step based on the highest similarity between its weight vector and the input.
\end{itemize}

\subsubsection{Reinforcement Learning (RL)}~\par

Reinforcement Learning (RL) is an ML paradigm designed to teach an agent (i.e., the learning entity) to make local decisions and take actions to maximize a cumulative long-term reward (via feedback from the environment) \cite{xie19}. In contrast to SL and UL, RL places a substantial emphasis on the temporal aspect. Consequently, the error metric in RL is distributed over time \cite{amin21}.
 
\begin{figure}[htbp]
\centerline{\includegraphics[scale=0.5]{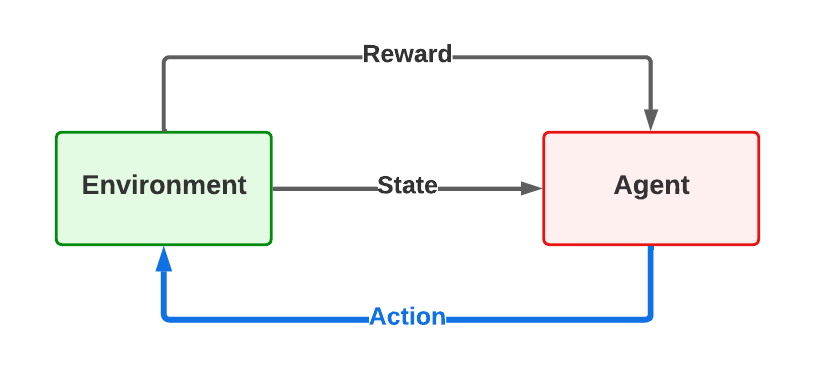}}
\caption{RL Typical Scenario}
\label{rlscen}
\end{figure}

As shown in Figure \ref{rlscen}, the agent monitors the state of the environment, and at each time step chooses an action, receives an immediate reward indicating how good or bad the action is, and transitions to the next state. The agent's goal is to learn the optimal behavioral policy to maximize the expected long-term reward.

The policy (i.e., the appropriate action for the current state) is a combination of exploration and exploitation. The agent can use exploitation to re-select the action that gave the highest reward in the current state, or can use exploration to select a different action in anticipation of a better reward \cite{ku17}.

In RL, a value function is used to compute the long-term reward of an action given a state.  One of the most renowned value functions is the Q-function, utilized by Q-learning to learn a table that contains all state-action pairs and their respective long-term rewards \cite{xie19}, and is responsible for assessing the quality of an action taken in any given state within the system \cite{ouhab20}. Q-learning is a model-free RL method to teach an agent how to act according to the state and the observations from the environment. Today, it is the basis for existing RL models.

Deep Reinforcement Learning (DRL)  constitutes a subset of RL that leverages the capabilities of DNNs in conjunction with RL models, exploiting the powerful function approximation property of DNNs. Given a state-action pair as input, DRL excels at estimating the long-term reward associated with it. This estimation, in turn, guides the agent in selecting the most favorable action. DRL effectively addresses some of the limitations encountered in traditional RL (e.g., low convergence rate to the optimal action policy, inability to solve problems with high-dimensional state and action spaces).

As RL evolved into DRL, Q-learning evolved into Deep Q-learning \cite{amin21} by replacing the Markov Decision Support (MDS) framework with DNN, thus solving the problem of multiple states and massive data.

\subsubsection{Federated Learning (FL)}~\par

As an alternative to conventional ML techniques, Federated Learning (FL) is one of the most attractive techniques that allows a heterogeneous set of devices to train an ML model without sharing their raw data \citep{guerra23} \citep{zheng23} \citep{nguyen21}. While improving privacy and communication efficiency, FL also leverages massive, distributed data and computational resources in constrained networks (e.g., IoT networks) \citep{zhao23}. For example, in FL, only learning model updates are transmitted between end devices and the FL aggregation server \citep{khan21}.

\begin{figure}[htbp]
\centerline{\includegraphics[scale=0.5]{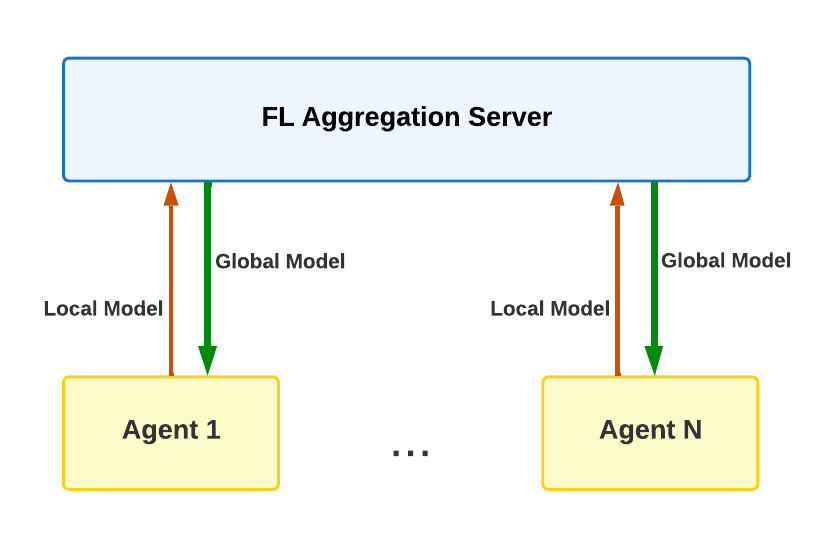}}
\caption{FL Typical Scenario}
\label{flscen}
\end{figure}

The FL process generally consists of three steps, as shown in Figure \ref{flscen} \cite{imteaj22} \cite{pervej23}: Training task and global model initiation (initially, the central server determines the task requirement and the target application), local model update (each participant trains the model using its local data), and global aggregation (when it receives the local models from the participants, the FL aggregation server performs the aggregation and generates the updated global model).

FL algorithms can be classified according to two primary dimensions, namely data partitioning and network structure \cite{nguyen21}. In terms of data partitioning, FL can be divided into three categories:  Horizontal FL (HFL), where all learning clients collaboratively train a global FL model using local datasets with consistent feature spaces but varying sample spaces, Vertical FL (VFL), where clients share identical sample spaces but differ in feature spaces, and Federated Transfer Learning (FTL) where clients have both distinct sample spaces and feature spaces. Considering the network structure, FL can be divided into two categories: Centralized FL (CFL), characterized by a central server coordinating a group of clients for model training, and Decentralized FL (DFL), where all clients are interconnected in a peer-to-peer (P2P) manner to collectively perform model training.

Note that FL can be combined with RL to form Federated Reinforcement Learning (FRL) \cite{qi21}. As a new and potential branch of RL, FRL can make learning safer and more efficient while taking advantage of FL. Similarly, DRL and FL can be combined to form Federated Deep Reinforcement Learning (FDRL).

\textcolor{blue}{In summary, SL algorithms are generally used to perform classification and regression tasks, while UL and RL algorithms are more suitable for clustering and decision-related tasks, respectively \cite{xie19}. The choice of the most suitable machine learning method primarily depends on factors such as the dataset nature, size, and the specific problem at hand. However, the application of ML techniques introduces several challenges that should be considered. For example, training ML models often involves transferring data from its sources, such as IoT devices, to a central system. This approach faces several problems: communication overhead, latency, energy consumption, and privacy concerns \cite{guerra23}.}

It should be noted that there is another type of learning, which is semi-supervised learning.  It is a type of learning where the algorithm has to deal with a part of the training dataset that is labeled, while another part of this data is not labeled beforehand \citep{alonso21}. One of the simplest and most efficient techniques is pseudo-labeling \citep{xie19}, where the labeled data is used to train a model via SL techniques, and then the trained model is used to predict pseudo-labels of the unlabeled data. Finally, all data are combined to train the model again.

\section{Intelligent Network Softwarization for IoT}
As previously discussed, the large number of heterogeneous devices in IoT networks poses enormous challenges. This opens up many research areas to build dynamic IoT networks and overcome the limitations. Network softwarization and ML techniques are enabling technologies that should be considered to address the above issues. In this section, we present the advantages of using SDN, NFV, and ML in IoT networks. First, we present the existing applications of network softwarization techniques in IoT. Then, we highlight the existing work on intelligent network softwarization. Finally, we review the state of the art in combining ML, SDN, and NFV with IoT networks.

\subsection{Network softwarization for IoT}
In IoT, network scalability and flexibility are critical. SDN and NFV are suitable network softwarization technologies to enable these functionalities. This topic has received a lot of attention from the research community. In the following, we present some relevant papers on this topic.

Considering the major issue of energy consumption in IoT networks, the authors in \citep{shuker19} propose a novel SDN-based routing technique for IoT that allows different devices and resources to connect to each other with minimum energy consumption. Necessary measurements have been performed and the power consumed in each sensor has been measured. The authors simulate the proposed work with \textit{NS2} network simulator and present a performance analysis (simulation parameters are mentioned). The metrics used are throughput and Packet Delivery Ratio (PDR). However, even though the authors assume that their solution improves the network performance, no comparison with other solutions (e.g., the commonly used RPL routing protocol) is made.

\textcolor{blue}{In \citep{saha20}, the authors propose an energy-aware SDN- and NFV-enabled architecture for IoT. SDN and NFV provide a powerful framework for optimizing data aggregation in IoT networks by enabling dynamic resource allocation to ensure scalability and increase efficiency in handling the diverse and dynamic data generated by IoT devices.} The authors argue that data aggregation (which can be defined as an NFV instance and dynamically deployed to the required IoT nodes) can further reduce network traffic and improve the overall resource utilization. The solution makes appropriate changes and optimizations to various network functions to cope with IoT networks. After detailing the proposal, the authors proceed to a simulation using \textit{Cooja}. Extensive evaluation confirms that the proposed solution outperforms its counterparts in terms of energy consumption and PDR. It should be noted that the proposed solution is based on the \(\mu\)\textit{SDN} as proposed by authors in \citep{baddeley18}.

\textcolor{blue}{In \cite{yazdinejad20}, the authors explore the potential of merging blockchain and SDN to address certain challenges in IoT networks. They propose a secure and energy-efficient blockchain-enabled architecture of SDN controllers for IoT networks. The architecture includes a cluster structure and introduces a novel routing protocol. While SDN optimizes routing through centralized control, the use of blockchain enables high resilience to attacks (given its decentralized nature) and enhances the security and accountability of SDN routing decisions. Through experimentation, the authors demonstrate that their proposed routing protocol outperforms existing counterparts in terms of performance and energy efficiency.  We note that the authors omit the well-known RPL protocol when comparing the proposal with its counterparts.}

In \citep{zarca20}, the authors introduce the concept of virtual IoT honeynets as a strategy for countering cyberattacks in softwarized IoT networks. Enhanced with SDN and NFV support, honeynets operate as virtualized services, replicating real IoT network configurations to divert potential attackers away from actual targets. The authors proceed to implement and assess their solution, showcasing its effectiveness in mitigating cyberattacks. The deployment of these honeynets relies on user-defined security policies. As future research, we suggest to consider ML techniques and especially RL to make the system automatically define and improve its security policies.

In \citep{molinazarca20}, the authors introduce a security orchestration framework that is both semantic-aware and policy-driven. This framework is designed for orchestrating security measures autonomously in softwarized IoT networks, with a specific focus on ensuring conflict-free security orchestration while optimizing the allocation of security VNFs. The proposal has been validated in a real testbed, demonstrating its feasibility and performance in detecting conflicts. As future work, we propose to consider incorporating intelligence into the solution to enable self-driven security orchestration.

The authors in \citep{haque21} design an SDN- and NFV-enabled IoT network architecture that considers resource and energy constraints as a top priority. They extensively evaluate the performance of their solution using \textit{Cooja}, and the solution outperforms its counterparts in terms of PDR, network lifetime, and overhead reduction, while being compatible with standard protocols. This paper extends the previously cited reference \citep{saha20} by presenting a novel optimization model. We are pleased to note that the authors provide the prototype implementation code for future reproduction and extension, a practice that we strongly encourage to stimulate research efforts in this area. As future work, they suggest considering a real testbed implementation for a better evaluation of the proposal. What we suggest as a future direction is to consider ML techniques to propose an intelligent solution of the optimization problem.

In \citep{hajian22}, the authors propose a novel mechanism for load balancing routing and virtualization in IoT networks using SDN. The authors recall that load balancing has four main goals: resource utilization, QoS, resilience, and scalability. Their approach leverages the \textit{OpenFlow} protocol to directly monitor link load information and network operational status, allowing for load balancing routing decisions to be tailored to individual flows across various IoT applications. After introducing the proposed algorithm, the authors evaluate its efficiency through simulation using \textit{NS2}. It has been compared with existing algorithms, and the simulation results support the findings. The simulation was performed for different use cases with different requirements. Note that this solution assumes a static network topology without mobile nodes, while real-world IoT networks often feature dynamic elements, including mobile nodes.

In \citep{shamsan22}, the authors propose a novel architecture to overcome the limitations posed by resource constraints in IoT through softwarization. The authors note that while existing works on softwarization within the IoT realm have tended to be specialized, focusing on solving specific problems, their proposed architecture takes a broader approach. It is designed to be a versatile solution aligning with the standard architectures of SDN, NFV, and IoT. The proposal is built and simulated using the \textit{Mininet} network emulator. The experimental outcomes show that there is a significant improvement in the performance of the proposed architecture over a conventional IoT system.

In \citep{jaadouni22}, the authors investigate various SDN- and NFV-based architectures tailored for edge-cloud-oriented IoT environments. Their primary objective is to assess the feasibility of deploying, administering, and disseminating network services to end-users within these contexts. They highlight that, so far, challenges linked to communication between edge and cloud components and the provision of network services in IoT networks have not been adequately addressed. However, no simulation of the architectures was conducted, making it difficult to consider the effectiveness of the proposal.

\textcolor{blue}{In conclusion, we can state that there is currently a great effort to fully softwarize IoT networks through SDN and NFV technologies, which is confirmed by the reviewed papers, in order to address the main challenges of IoT. The integration of SDN in IoT networks offers many advantages, thanks to the centralized global control and decision making it provides \citep{shamsan22}. In particular, it is used to optimize routing in terms of QoS performance and energy efficiency \citep{yazdinejad20} \citep{shuker19} \citep{baddeley18} and provides better results than traditional constrained routing protocols such as RPL, as it acts based on a global visibility of the network status. Routing protocols can be easily extended to provide load balancing, which can ensure better energy conservation of the nodes \citep{hajian22}. It is also an enabler to perform data aggregation, thus reducing network traffic and improving resource utilization \citep{saha20} \citep{haque21}. It is important to recall that the combination with NFV enhances these benefits, especially for security purposes. Both technologies make it possible to achieve high IoT network security by facilitating the setting of user-defined security policies and global policies, and ensuring the efficient deployment of network security services \citep{zarca20} \citep{molinazarca20} \citep{jaadouni22}.}

\textcolor{blue}{However, we point out that these reviewed references do not include intelligence in the solution. Intelligence, via ML techniques, is certainly a key feature to ensure an efficient network softwarization for IoT. It makes it possible to address the challenges posed by SDN and NFV. In particular, since SDN enables dynamic network configuration, manual optimization of network parameters can be challenging (especially for large-scale heterogeneous IoT networks), so ML can, for example, perform real-time network performance data collection and automatically optimize parameters. Another pressing issue is that SDN and NFV introduce new attack vectors, and security policies may need to be constantly updated to address evolving threats. ML can be used to identify unusual patterns in network behavior that may indicate security threats. It enables the system to learn from historical data and adapt to new and emerging cyber threats without the need for frequent manual updates.}

\subsection{ML-based network softwarization}
Nowadays, ML techniques are a driving force of several domains, and it is certainly a booster for network softwarization. It can potentially be used to address many challenges in networking, including design, implementation, performance, and verification. Therefore, it is quite normal that academia has focused on the topic of enabling ML for SDN and NFV in recent years, as presented below.

In \citep{ku17}, the authors propose an RL-based algorithm for selecting an appropriate path for a Service Function Chain (SFC - i.e., a sequential chain of VNFs) in an SDN- and NFV-enabled network. The proposed solution selects an appropriate path depending on the network conditions, thus ensuring an efficient service chaining environment. The selection method is based on Q-learning. The authors chose RL techniques because they believe that it is the most appropriate ML technique for decision making problems.  The reward depends on computational and bandwidth usage.  The authors implement a simulator with \textit{Java} and evaluate the performance of their proposal in comparison with the greedy method. The solution outperforms its counterpart by taking network conditions into account. However, we note that the authors did not consider constrained networks (e.g., IoT networks) and thus did not consider an energy-related metric to evaluate energy efficiency.

\textcolor{blue}{The authors in \citep{le18} propose the application of emerging technologies such as big data, ML, SDN, and NFV to perform traffic clustering (important for self-organization), forecasting (important for developing powerful optimization applications), and management in 5G networks to improve the efficiency and quality of mobile networks. The authors build a powerful practical framework, which is evaluated by applying it to a real dataset. We note that the peculiarities of IoT networks are not considered in this work, even though IoT networks share many characteristics with 5G networks (e.g., heterogeneity, exponential growth, etc.). We also note that authors have mainly focused on SL and UL techniques (e.g., \textit{k}-means, linear regression, neural networks, etc.) and have not considered RL-based solutions.}

\textcolor{blue}{In \citep{troia19}, the authors propose a novel planning and provisioning solution for SDN- and NFV-enabled metropolitan networks based on ML. The proposed framework achieves dynamism and adaptability (i.e., guarantees fair behavior towards past, current, and future requests). Network planning aims to optimize network resources both offline (i.e., network capacity planning) and online (i.e., resource allocation). The authors recall that previous work in this area has shown that ML ensures a network configuration that is faster than integer linear programming optimization and that is more accurate than heuristic solutions. We note that the evaluation is in its early stages. More extensive experiments and tests are required to determine the effectiveness of the solution in real-world scenarios.}

\textcolor{blue}{The authors in \citep{jiang18} propose an SDN- and NFV-based proof-of-concept testbed for ML-enabled network management. The authors recall that unlike the physical layer, theoretical analysis and numerical simulations are generally not rigorous for the management layer, where ML is applied to realize intelligent and self-organized networks. By proposing this solution, the authors aim to encourage innovative work in the field of intelligent network softwarization. Through experiments, the effectiveness of ML-based network management and the flexibility of network softwarization techniques are demonstrated. A derivative of this testbed can be developed to take into account the peculiarities of IoT networks.}

\textcolor{blue}{In \citep{abdulqadder20}, the authors propose a multi-layered intrusion detection and prevention solution for SDN- and NFV-enabled cloud of 5G networks. The proposed approach defends against security attacks based on various intelligence techniques (e.g., game theory, SOM, DRL, etc.). The proposed system is evaluated against various security attacks (e.g., spoofing, Distributed Denial of Service (DDoS), host location hijacking, etc.), and the novel mechanism has proven to be effective in detecting and preventing such attacks. As future work, the authors plan to consider blockchain, which is a very insightful technology for security enhancement when integrated to the SDN control plane. We suggest considering these advances in security for vulnerable IoT networks.}

\textcolor{blue}{In \citep{baranda20}, the authors propose a solution to achieve network service autoscaling for a 5G end-to-end service platform that integrates AI and ML techniques for all decision processes of the MANO stack. The authors perform the evaluation of their proposal, and the results prove its effectiveness. This is a first prototype that integrates closed-loop ML-based decision making for 5G network management. In this work, the authors apply the solution to autoscaling, while noting that it could also be adapted to any automated management decision process (e.g., intelligent resource management at the resource layer). We support the authors' choice of RL as the ML technique to be adopted for this purpose.}

\textcolor{blue}{The authors in \citep{ilievski21} study the use of SL in NFV environments for network traffic classification. They investigate the performance of six different SL techniques. The analysis performed concludes that the decision tree algorithm is the most suitable for performing classification-related tasks in NFV environments, as it induces less delay while being reliable. At this point, we note that the authors do not consider many packet-related details for classification purposes, such as packet payload. Authors recall that NFV is a true enabler of 5G networks. However, they do not consider SDN, which is not covered in this work yet.}

\textcolor{blue}{In \citep{nouruzi22}, the authors present a DRL-based framework tailored for online end-user service provisioning in an NFV-enabled network. They pose an optimization problem with the goal of minimizing the cost of network resource utilization, and introduce a Deep Q-network as a solution to this optimization challenge. In particular, the authors provide an assessment of the computational complexity associated with their proposed approach. Furthermore, to measure the performance of their framework, the authors benchmark it against various baseline methods. The results, derived from various parameter settings, show the effectiveness of the framework}.

\textcolor{blue}{In \citep{ramya23}, the authors propose a novel mechanism for traffic-aware dynamic controller placement in SDN using NFV. An ML approach (based on RNN) is developed for network traffic management by predicting the number of controllers to be placed in the network. The prediction mechanism is developed as a VNF, which is then deployed. This solution effectively combines SDN, NFV, and ML to achieve network automation. Experimental evaluation validates the effectiveness of the proposal, which can be extended to consider reliability.}

\textcolor{blue}{The authors in \citep{yungaicela23} propose an SDN- and NFV-based framework for autonomous defense against slow-rate DDoS attacks. Unlike traditional DDoS attacks, which flood the target with a large amount of traffic in a short period of time, slow-rate DDoS attacks are more subtle and prolonged. The solution uses DL to detect attacks and RL to mitigate them. An NFV-powered moving target defense mechanism is included to increase effectiveness and flexibility. It dynamically configures the network to puzzle the attack surfaces. Authors perform extensive simulations to evaluate the effectiveness of the solution, which shows its high ability to perform optimal DDoS mitigation scenarios while ensuring high adaptability. The authors share the prototype source code, which is very beneficial for other researchers to build on future improvements.}

\textcolor{blue}{In summary, many research efforts have been made to merge ML with network softwarization, especially SDN and NFV. Intelligence enables efficient NFV orchestration by reducing computation and bandwidth consumption \citep{ku17}, and optimized NFV service provisioning by minimizing the cost of network resource utilization \citep{nouruzi22}. The placement of NFV functions is made more dynamic and traffic-aware \citep{ramya23}. Intelligence also enables more efficient softwarized network management and planning \citep{troia19} \citep{jiang18}. The combination of SDN, NFV, and ML is a booster for 5G networks and beyond, enabling performance improvements through intelligent traffic classification and management \citep{le18} \citep{ilievski21} and service autoscaling \citep{baranda20}. The result for security enhancement is very positive \citep{abdulqadder20} \citep{yungaicela23}, as intelligence enables dynamic autonomous defense against various types of attacks and intrusions.}

\textcolor{blue}{However, we note that authors usually consider SDN as the main key enabler for network softwarization and omit NFV. Some interesting ML techniques, such as FL, are not considered by many of the reviewed works, even though they are highly relevant in large-scale distributed networks where nodes have the computational capability to perform part of the learning process. We share the concern reported in many papers regarding the unavailability of network datasets and the need to create high-quality standardized datasets \citep{wang18} \citep{mestres17}. We also note that even if some works consider different networks, they omit IoT networks, which are continuously growing and require efficient intelligent network softwarization techniques.}

\subsection{Intelligent IoT network softwarization}
ML plays an essential role in creating smarter IoT networks, as it has shown remarkable results in various domains. It allows IoT networks to learn from experience to make them more robust against vulnerabilities and failures, and to improve performance, thus addressing the major challenges. As the architecture of network softwarization enables the integration of ML, we strongly believe that we will soon see a fusion of ML and network softwarization techniques with IoT. In the following, we present the most recent and relevant works, even though there is a small number of publications on the topic.

In \citep{ouhab20}, the authors present an energy-efficient clustering and routing algorithm tailored for monitoring large-scale IoT networks based on SDN technology. They present a novel modeling approach structured around a dual-level control mechanism to address the challenge of the lack of an efficient routing protocol capable of handling a significant number of devices while ensuring low-power data forwarding. The authors acknowledge that current algorithms are poorly suited for efficient routing in large networks. These algorithms do not effectively scale to large device populations, nor are they energy efficient, especially in scenarios involving mobile devices. To address these problems, the authors employ a two-tiered approach: at the first tier, they implement a multi-hop clustering routing mechanism in conjunction with RPL routing, while at the second tier, they employ an SDN controller in combination with a Q-routing algorithm based on Q-learning. This two-tier setup enables intelligent network management on a global scale. To evaluate the effectiveness of their model, the authors perform simulations using the \textit{Cooja} platform. The results clearly show that their proposed model delivers significantly better results compared to current state-of-the-art solutions.

\textcolor{blue}{The authors in \citep{lin20} propose a novel system architecture and design of an AI-enabled, softwarized 5G and IoT network. This combination provides agility and flexibility in resource placement and utilization, as softwarization in 5G and IoT enables more programmable and dynamic network configurations. The authors recall that a variety of industries have been made intelligent using AI, ML, and DL, so the proposed architecture aims to leverage the advances in AI technologies to improve the efficiency of 5G and IoT networks. The architecture is presented and implementation challenges are listed (e.g., technical complexity, availability of open source resources, etc.), but few details are provided to enable implementation.}

In \citep{sellami20}, the  authors propose a DRL approach for energy-efficient task assignment and scheduling in an SDN-based fog IoT network. Task scheduling is a common problem that often involves complex online decision making. The primary goal of this approach is to minimize network latency while ensuring optimal energy efficiency. By integrating DRL, the method leverages intelligent agents capable of learning and making better decisions through direct interaction with the environment. The authors use a testbed setup to evaluate and analyze the performance of their proposed approach. The evaluation results show its superiority over both deterministic and random task scheduling strategies. However, it is worth noting that certain data-related challenges, such as privacy concerns, need to be addressed, and FL emerges as a promising enabling technology to effectively address these issues.

In \citep{memos22}, the authors highlight the significant problem posed by the uncontrolled proliferation of insecure IoT-based devices. They outline the vulnerabilities associated with these devices and propose a solution using an NFV infrastructure in conjunction with emerging technologies to enable intelligent management and enhanced protection against botnet attacks. Intrusion attacks, particularly in the form of botnets, are a common threat in IoT networks. Botnets consist of a large network of compromised devices acting as bots and are often associated with DDoS attacks, Man in the Middle (MitM) attacks, malware distribution, and other security risks. IoT botnets are of particular concern due to their rapid propagation, which allows them to have a greater impact than traditional botnets. The authors present a potential NFV architecture that incorporates several emerging technologies into the solution. These technologies include a virtual honeynet that acts as a decoy to trap attackers, cloud computing to reduce latency in providing protection against vulnerabilities, and ML to analyze cyber-threats. While the authors provide a detailed description of their proposed solution, it is important to note that it has not been implemented or evaluated for performance, particularly in real-world scenarios involving known and zero-day attacks.

In \citep{ros22}, the authors present a novel approach involving a modified DRL agent designed for dynamic resource placement within IoT network slicing. This specialized agent engages with controllers and orchestrators to effectively manage the installation of flow rules and the allocation of physical resources in NFVI. The approach introduces a unique formulation that incorporates completion time and criticality criteria. To evaluate the proposed method, the authors conduct simulations using the \textit{Mininet} emulator. They assess performance using key metrics such as delay and PDR. The comparative analysis involves three agents: the proposed modified DRL agent, an unmodified DRL agent, and an experience-based allocation agent. The results of the evaluation demonstrate that the proposed solution surpasses its counterparts in terms of performance.

While most existing works assume that services are represented as SFCs, which are chains (for reference, we previously reviewed the matter in \cite{ku17}), the authors in \cite{xie22} consider that network services in IoT networks exhibit greater complexity and diversity. Consequently, they propose a more suitable representation known as VNF Forwarding Graphs (VNF-FGs), which are essentially Directed Acyclic Graphs (DAGs). They point out that prior research has failed to fully leverage this specific graph structure, rendering them suboptimal or unsuitable for IoT networks. In light of this, the authors delve into the VNF-FG placement problem within dynamic IoT networks. To effectively exploit the graph structures of services and address the challenges of dynamic IoT networks, they employ a combination of Graph Neural Network (GNN) and DRL. Their resulting algorithm, named \textit{Kolin} is designed for efficiency. Extensive simulation results suggest that \textit{Kolin} outperforms state-of-the-art solutions (e.g., First Fit Dijkstra, Greedy) in terms of system cost, acceptance rate, and computational complexity. 

In \citep{samadi22}, the authors study an ML routing protocol in mobile SDN-based IoT. The authors note that RPL, one of the prominent routing protocols in IoT, assumes that the network has no movement, and all nodes are static. Therefore, this routing protocol does not provide a mechanism to support mobile nodes. To overcome this problem, they propose to combine ML and SDN. Unfortunately, the solution is not detailed, and no simulation is performed to evaluate the performance based on relevant metrics.

To address edge FL challenges in large-scale heterogeneous IoT networks, the authors in \cite{tam22} introduce a model that integrates SDN and NFV. This integration facilitates the deployment of NFV-enabled edge FL aggregation servers, enhancing automation and control. The proposed solution is powered by ML, specifically utilizing Multi-Agent Deep Q-Networks (MADQNs) to enable self-learning through softwarization. The authors conducted simulations to evaluate the performance of their solution, and the results demonstrate that it outperforms reference methods in terms of QoS metrics.

In \cite{keshari23}, the authors introduce a smart and energy-efficient traffic flow control strategy for SDN-based IoT networks. Their proposal aims to identify the optimal set of boundary nodes for each cluster within the network, which helps reduce the total number of potential paths between clusters. The selection of these optimal border nodes is based on criteria involving maximum energy and minimum distances. To identify this set of optimal border nodes, the authors employ the Lion Swarm Optimization Algorithm (LSOA), a nature-inspired optimization technique. Although LSOA is not a machine learning method, it is a recent and noteworthy approach in this field. The simulation results conducted using \textit{MATLAB} indicate that border nodes selected using LSOA outperform those selected using other state-of-the-art metaheuristic algorithms. This leads to enhanced network efficiency through energy conservation. As a future research direction, ML algorithms could be considered.

In \cite{deoliveira23}, the authors introduced an ML approach to place security VNFs (e.g., firewall) to mitigate DDoS attacks on industrial IoT. The authors note that few works have used NFV to detect and mitigate threats on industrial IoT networks, and even fewer have considered network performance indicators when placing VNFs. The proposed approach is novel as it considers NFV performance indicators as decision variables (e.g., deployment time, computation resource utilization, and memory consumption). Experiments show that ML is an effective alternative, as it shows 99.40\% accuracy with respect to ideal placement. The authors have considered and compared several ML techniques (e.g., \textit{k}-NN, RF). It is good to note that the authors made all the produced code and data publicly available, which allows the community to reproduce the experiments. We note that the authors considered only one specific attack, but their approach could be applied to mitigate other types of attacks, such as MitM attacks.

\textcolor{blue}{In summary, we note that efforts to merge ML, network softwarization (mainly SDN and NFV), and IoT have been increasing recently \citep{lin20}. The combination of ML algorithms and network softwarization enables intelligent routing for cluster-based IoT networks. These algorithms process the data centralized by the SDN controllers to form and manage clusters \citep{keshari23}, and perform optimized routing, mainly based on Q-learning RL technique, which adjusts routing decisions to meet energy consumption constraints and QoS requirements \citep{ouhab20} \citep{samadi22}. It also enables intelligent VNF management with consideration of system cost (mainly using DRL), which enables self-driven network service orchestration for complex large-scale IoT NFV-enabled networks \citep{xie22} \citep{ros22}. Security is a major concern in IoT networks, and as softwarization opens many attack vectors, the use of ML allows to counteract them \citep{memos22}. As security VNFs are deployed in the IoT network, intelligence allows autonomous placement and orchestration, ensuring high adaptability to requirements that may vary, and thus ensuring the dynamism of VNF deployment \citep{deoliveira23}. It is also interesting to note the combination of all the above mentioned technologies with other emerging softwarization technologies, namely edge and fog computing, where intelligent network softwarization enables optimized task assignment and scheduling in IoT fog networks, with the aim of minimizing network latency and balancing loads while ensuring energy efficiency \citep{sellami20}. Other issues related to data filtering and reduction are addressed through the use of ML, where IoT systems are enabled to decide which data is unnecessary and could be filtered at the edge level to conserve network bandwidth \citep{tam22}.}

\textcolor{blue}{While the integration of intelligent IoT network softwarization has gained momentum (the number of research papers on intelligent IoT network softwarization mainly increased from 2022), several challenges remain and require further attention and research to achieve effective solutions. Among them, we recall the security and privacy concerns as IoT devices generate and exchange sensitive data. The softwarization introduces new attack vectors, and protecting against cyber-threats, data breaches, and ensuring user privacy remains a significant challenge. We also recall the interoperability issues, as smart softwarization solutions must seamlessly integrate with a variety of devices, protocols, and communication standards to ensure a cohesive and efficient IoT ecosystem. As the IoT intelligent softwarization landscape continues to evolve, researchers will have to work toward comprehensive solutions that balance efficiency, security, and usability.}

\section{Discussion}
It is evident that academia is fully aware of the potential of SDN, NFV, ML, and IoT as key enabling technologies. Each technology has received a lot of attention on its own, and as a result, recent research papers have proposed to merge them. Since network softwarization is essential to ensure efficient, agile IoT networks, and since ML algorithms are more appropriate for softwarized networks than traditional ones, we are moving toward intelligent network softwarization for IoT very soon.

\textcolor{blue}{In Table \ref{tb:sum}, we summarize the recent advances in the field of intelligent network softwarization for IoT. The examination of these works reveals that those that exploit the full potential of the aforesaid technologies in combination are still sparse. There are important advances in intelligent network softwarization, but they are not always applicable to IoT constrained networks. Many works that consider network softwarization only consider SDN and omit NFV, even though the combination of both ensures the efficiency of the softwarization.}

\textcolor{blue}{Recent advances have shown that network softwarization, especially when combined with ML, is a great solution to address key IoT challenges. Since network softwarization enables network programmability and centralized control, it is a booster for routing in such constrained networks. Routing decisions are dynamically adjusted to achieve the best QoS performance (e.g., low latency, high throughput, etc.) while ensuring the lowest possible energy consumption. This is made possible by the integration of ML, especially RL and DRL. Routing protocols can be easily extended to provide load balancing for even better energy savings. Furthermore, when data aggregation is enabled, network traffic is reduced and resource utilization is improved.}

\begin{longtblr}[
   label = tb:sum,
  caption = {Recent Advances on the Combination of Network Softwarization, ML, and IoT},
]{
  width = \linewidth,
  colspec = {Q[71]Q[62]Q[62]Q[113]Q[67]Q[65]Q[100]Q[390]},
  row{1} = {c},
  row{2} = {c},
  row{14} = {fg=blue},
  row{15} = {fg=blue},
  row{16} = {fg=blue},
  row{17} = {fg=blue},
  row{18} = {fg=blue},
  row{19} = {fg=blue},
  row{21} = {fg=blue},
  row{22} = {fg=blue},
  cell{1}{1} = {r=2}{},
  cell{1}{2} = {c=2}{0.124\linewidth},
  cell{1}{4} = {r=2}{},
  cell{1}{5} = {r=2}{},
  cell{1}{6} = {r=2}{},
  cell{1}{7} = {r=2}{fg=blue},
  cell{1}{8} = {r=2}{},
  cell{3}{7} = {fg=blue},
  cell{4}{7} = {fg=blue},
  cell{5}{7} = {fg=blue},
  cell{6}{7} = {fg=blue},
  cell{7}{7} = {fg=blue},
  cell{8}{7} = {fg=blue},
  cell{9}{7} = {fg=blue},
  cell{10}{7} = {fg=blue},
  cell{11}{7} = {fg=blue},
  cell{12}{7} = {fg=blue},
  cell{13}{7} = {fg=blue},
  cell{20}{7} = {fg=blue},
  cell{23}{7} = {fg=blue},
  cell{24}{7} = {fg=blue},
  cell{25}{7} = {fg=blue},
  cell{26}{7} = {fg=blue},
  cell{27}{7} = {fg=blue},
  cell{28}{7} = {fg=blue},
  cell{29}{7} = {fg=blue},
  cell{30}{7} = {fg=blue},
  cell{31}{7} = {fg=blue},
  cell{32}{7} = {fg=blue},
  hline{1,3-33} = {-}{},
  hline{2} = {2-3}{},
}
\textbf{Ref.} & \textbf{Network Softwarization} &  & \textbf{ML Technique(s)} & \textbf{IoT Networks} & \textbf{~Evaluation} & \textbf{Applica- tion case(s)} & \textbf{Objective(s)}\\
 & \textbf{SDN} & \textbf{NFV} &  &  &  &  & \\
\citep{baddeley18} & \cmark & \xmark & \xmark & \cmark & \cmark & Network Architecture & Evolving SDN to be adapted to constrained IoT networks.\\
\citep{shuker19} & \cmark & \xmark & \xmark & \cmark & \cmark & Routing & Improving routing protocols for IoT networks.\\
\citep{saha20} & \cmark & \cmark & \xmark & \cmark & \cmark & Network Architecture & Providing an energy aware softwarized architecture for IoT.\\
\citep{yazdinejad20} & \cmark & \xmark & \xmark & \cmark & \cmark & Network Architecture, Security & Proposing an energy efficient SDN controller architecture for IoT networks with blockchain-based security.\\
\citep{zarca20} & \cmark & \cmark & \xmark & \cmark & \cmark & Security & Proposing virtual honeynets to mitigate cyberattacks in SDN- and NFV-enabled IoT networks.\\
\citep{molinazarca20} & \cmark & \cmark & \xmark & \cmark & \cmark & Security & Proposing a semantic-aware security orchestration framework in softwarized IoT systems.\\
\citep{haque21} & \cmark & \cmark & \xmark & \cmark & \cmark & Network Architecture & Proposing a resource-aware SDN and NFV-based IoT network architecture.\\
\citep{hajian22} & \cmark & \xmark & \xmark & \cmark & \cmark & Routing & Proposing a mechanism for load balancing routing and virtualization for SDN-enabled IoT.\\
\citep{shamsan22} & \cmark & \cmark & \xmark & \cmark & \cmark & Network Architecture & Proposing a novel IoT architecture based on network softwarization.\\
\citep{jaadouni22} & \cmark & \cmark & \xmark & \cmark & \xmark & Network Architecture & Presenting SDN- and NFV-based architectures for edge-cloud oriented architectures.\\
\citep{ku17} & \cmark & \cmark & RL & \xmark & \cmark & VNF Placement & Studying RL-based SFC path selection in softwarized networks.\\
\citep{le18} & \cmark & \cmark & SL, UL & \xmark & \cmark & Traffic Optimization & Proposing a novel framework for 5G traffic clustering, forecasting, and management based on big data, ML, SDN, and NFV.\\
\citep{troia19}  & \cmark & \cmark & \cmark & \xmark & \cmark & Network Management & Proposing an ML-based planning and provisioning tool for SDN- and NFV-enabled metropolitan networks.\\
\citep{jiang18}  & \cmark & \cmark & \cmark & \xmark & \cmark & Perform- ance Evaluation & Proposing an SDN- and NFV-based testbed for ML-enabled network management.\\
\citep{abdulqadder20}  & \cmark & \cmark & \cmark & \xmark & \cmark & Security & Proposing a multi-layered intrusion detection and prevention solution for SDN- and NFV-enabled cloud of 5G networks.\\
\citep{baranda20}  & \cmark & \cmark & \cmark & \xmark & \cmark & Network Management & Proposing a solution to achieve network service autoscaling for a 5G platform.\\
\citep{ilievski21}  & \xmark & \cmark & SL & \xmark & \cmark & Traffic Optimization & Evaluating different SL techniques for traffic classification in NFV-enabled networks.\\
\citep{nouruzi22} & \xmark & \cmark & DRL & \xmark & \cmark & VNF Placement & Studying a DRL based framework for online service provisioning in NFV-enabled networks.\\
\citep{ramya23}  & \cmark & \cmark & RNN & \xmark & \cmark & VNF Placement & Proposing an intelligent mechanism for traffic-aware dynamic controller placement in SDN using NFV.\\
\citep{yungaicela23}  & \cmark & \cmark & DL, RL & \xmark & \cmark & Security & Proposing an intelligent SDN- and NFV-based framework for autonomous defense against slow-rate DDoS attacks.\\
\citep{ouhab20} & \cmark & \xmark & Q-learning & \cmark & \cmark & Routing & Providing an energy efficient clustering and routing algorithm for SDN-based IoT networks.\\
\citep{lin20} & \cmark & \cmark & \cmark & \cmark & \xmark & Network Architecture & Discussing the progress toward an AI-enabled SDN- and NFV-based 5G and IoT networks.\\
\citep{sellami20} & \cmark & \xmark & DRL & \cmark & \cmark & Network Management & Presenting a SDN-based dynamic task scheduling and resource management DRL approach for IoT.\\
\citep{memos22} & \xmark & \cmark & \cmark & \cmark & \xmark & Security & Proposing an NFV-based scheme for effective protection against bot attacks in AI-enabled IoT.\\
\citep{ros22} & \xmark & \cmark & DRL & \cmark & \cmark & VNF Placement & Proposing a modified DRL agent for dynamic resource placement in IoT network slicing.\\
\citep{xie22} & \cmark & \cmark & GNN, DRL & \cmark & \cmark & VNF Placement & Proposing a GNN-assisted DRL method for VNF-FG placing in softwarized IoT networks.\\
\citep{samadi22} & \cmark & \xmark & \cmark & \cmark & \xmark & Routing & Studying ML-based routing protocol in mobile SDN-enabled IoT networks.\\
\citep{tam22} & \cmark & \cmark & \cmark & \cmark & \cmark & Routing & Proposing a Multi-Agent Deep Q-Networks for efficient edge FL communications in SDN- and NFV-enabled IoT.\\
\citep{keshari23} & \cmark & \xmark & LSOA (not a ML technique) & \cmark & \cmark & Traffic Optimization & Proposing an intelligent energy efficient optimized approach to control the traffic flow in SDN-enabled IoT networks.\\
\citep{deoliveira23} & \xmark & \cmark & \cmark & \cmark & \cmark & Security & Introducing an ML approach to place security VNFs aiming an efficient mitigation of DDoS attacks on industrial IoT systems.
\end{longtblr}

\textcolor{blue}{It is also interesting to note that this combination of ML and network softwarization results in novel enabling architectures for IoT that take full advantage of NFV, where VNFs are autonomously and dynamically deployed based on real-time assessed needs. This traffic awareness leads to traffic optimization and service autoscaling capabilities. Network management is automated, enabling a self-driven network.}

\textcolor{blue}{The combination of these technologies is coupled with the opening of new attack vectors. The attention of academia to this critical issue is reflected in the number of works aimed at improving the security of IoT networks. Intelligent softwarized IoT networks can quickly and dynamically establish global policies and ensure the efficient deployment of network security services (e.g., firewalls, honeynets, etc.). The use of various ML techniques enables the identification of unusual patterns in network behavior that may indicate security threats. It enables the system to learn from historical data and adapt to new types of attacks and intrusions without the need for manual updates.}

\textcolor{blue}{Recent work also suggests the incorporation of other emerging technologies such as fog, edge, and cloud computing. Intelligent softwarized IoT networks can achieve optimized task assignment and scheduling in IoT fog networks, thus minimizing latency and balancing loads while ensuring energy efficiency. Using ML, IoT systems can decide which data is unnecessary and can be filtered at the edge to conserve network bandwidth.}

However, we point out that few works propose a consistent implementation and evaluation, due to lack of access to the necessary resources (e.g., network datasets, source codes). We join all the authors who strive to have high quality normalized network datasets (which can at least be generated through the implementation of testbeds or through network simulation) to be able to implement solutions and perform coherent performance evaluations. Having access to the code of the proposals and the simulation parameters is essential to be able to make comparisons and thus improve the proposals. Otherwise, research in this emerging area will be limited, but also scattered, as authors will not be able to improve and build on existing work. Having reproducible research is the key to real implementation in practice, to industry-based practical solutions (this may be possible if implementations are as close to real scenarios as possible).

It is undeniable that performing simulations to evaluate the proposal is very important for the development of this promising field, so the choice of the simulation platform is very important for the veracity of the results. In the early stages, the implementation can be done using a simulator, an emulator, or a testbed \citep{patel19}.  A simulator is the right choice in the first phase of the design because it provides a higher level of abstraction. Emulators map real-world devices to simulated ones, so the results are more reliable than those of simulators. Testbeds are much more accurate than software-based tools because they are a hardware-based tool for implementing and evaluating a solution.

\textcolor{blue}{We aim to have more publications on the integration of IoT with intelligent network softwarization. Obviously, the latter should be adapted to the peculiarities of IoT networks. We note that there are no works that define in detail an ML-enabled SDN- and NFV-based IoT network architecture, which is urgently needed as a baseline for further specific research. Following this, and after designing the solutions, the authors should consider the IoT constraints, such as energy constraints, when defining the evaluation metrics. It is important to make consistent comparisons, as we found that authors usually compare their proposals with conventional solutions, not with the smart counterparts.}

\textcolor{blue}{\section{Future Research Directions}}
 \textcolor{blue}{It is manifest that the scientific community is fully aware of the benefits offered by intelligent IoT network softwarization, and some researchers have even moved towards combining it with other emerging technologies. This leaves many areas open as interesting future research directions, a selection of which are presented in this section.}

 \textcolor{blue}{\subsection{Consideration of other emerging technologies}}

 \textcolor{blue}{Network softwarization is constantly evolving and incorporating more and more enabling technologies. One of these is Edge Computing (EC). It makes it possible to provide shorter service response times and reduce the cost of processing IoT data in the cloud, while ensuring the offloading of intensive computational tasks from less powerful devices to powerful edge servers. This increases the reliability and resilience of IoT systems and facilitates the move to a distributed architecture. EC also addresses security concerns by allowing sensitive data to be processed locally, reducing the need to transmit it over networks. Another paradigm to consider is fog computing, which complements EC by extending computing capabilities closer to the edge of the network.}

 \textcolor{blue}{Another technology to consider and which has gained significant attention with the advent of 5G is network slicing. It enables the creation of logically independent and customized virtual networks within a shared physical infrastructure. When combined with intelligent IoT network softwarization, it enables the creation of customized service offerings through isolation and service differentiation (i.e., each network slice can have its own performance characteristics, QoS policies, security measures, and resource allocation). It enhances security through the ability to define security perimeters and slice-specific security policies.}

 \textcolor{blue}{\subsection{Distributed intelligent IoT network softwarization}}

 \textcolor{blue}{As IoT networks grow exponentially and encompass billions of devices, there is an urgent need to move toward an architecture that avoids a Single Point of Failure (SPoF). This is possible by developing a distributed intelligent network softwarization architecture for IoT, where multiple controllers coexist in a clustered multi-tier scheme, ensuring scalability and reliability.  The integration of intelligent decision capabilities enables the network to predict potential failures, proactively mitigate risk, and optimize resource utilization across controllers.}

 \textcolor{blue}{Such a distributed architecture implies the possibility to propose distributed learning approaches. Federated Learning (FL) is a highly recommended technique to guarantee privacy, since it allows local model training without transmitting raw data to a central location. On the other hand, it enables collaborative training as the multiple learning entities work together to build a common model without exposing sensitive data. This maximizes resilience as FL mitigates the impact of individual node failures by allowing learning to continue on other operational nodes, ensuring robustness in distributed environments.}

  \textcolor{blue}{\subsection{Improvement of IoT security with blockchain}}

 \textcolor{blue}{In order to ensure real implementable solutions in daily life, security should be given the necessary attention as it is a cross-cutting aspect in the intelligent network softwarization of IoT. The integration of blockchain into softwarized IoT networks is an enabler to address security challenges. The distributed and immutable nature of blockchain provides an opportunity to enhance security, especially in managing access control, authentication, and authorization in SDN and NFV architectures.}
 
 \textcolor{blue}{Integrating blockchain into IoT systems requires addressing interoperability issues, as different blockchain platforms may not work together seamlessly. In addition, the performance overhead associated with blockchain consensus mechanisms could impact real-time responsiveness. Ensuring the security of blockchain transactions, while maintaining the performance and scalability requirements of SDN and NFV networks, remains a critical challenge. Overcoming these challenges requires the development of tailored consensus mechanisms, efficient data management strategies, and robust identity and access management solutions that are compatible with the dynamic and scalable nature of network softwarization.}

  \textcolor{blue}{\subsection{Standardization of IoT networking}}

 \textcolor{blue}{The challenge of standardization within softwarized IoT networks is critical to ensuring interoperability, seamless integration, and widespread adoption. The lack of consistent standards poses significant challenges. Divergent interpretations and proprietary implementations inhibit vendor collaboration, limit interoperability, and complicate multi-vendor deployments.}
 
 \textcolor{blue}{Establishing comprehensive and widely accepted standards is critical to fostering an ecosystem where diverse components from multiple vendors can seamlessly coexist, enabling innovation, reducing complexity, and ensuring a future-proof network infrastructure. Standardization efforts should focus on defining common protocols, data models, and interfaces to facilitate the consistent development, deployment, and management of IoT network softwarization solutions.}

~

\section{Conclusion}

\textcolor{blue}{IoT is nowadays a part of our daily lives, with millions of heterogeneous devices collecting valuable information through various sensors. Therefore, there is an important need to provide agile and scalable IoT networks. Considering the current emerging technologies, intelligent network softwarization (mainly SDN and NFV techniques) is certainly the enabler to address the IoT constraints and challenges, and meet the requirements in terms of QoS and energy efficiency.}

\textcolor{blue}{In this paper, we have analyzed the recent advances in the field of intelligent IoT network softwarization and concluded that more research efforts should be made towards implementable solutions that will positively impact our daily lives. To this end, many aspects such as standardization and security need to be fully addressed. Many emerging technologies, such as EC and blockchain, have demonstrated their effectiveness in various domains and may be the solution to these IoT networking matters when considered together with SDN, NFV, and ML.}

\textcolor{blue}{As future work, we plan to consider these technologies to propose an intelligent network softwarization architecture that addresses key IoT challenges. This architecture will take advantage of distributed control to incorporate intelligence techniques (mainly FL) to enable the transition to a self-configured, self-managed, and self-secure IoT. The general architecture is intended to be the basis for more specific work to optimize networks and enhance security in constrained IoT.}

\bibliographystyle{elsarticle-num-names} 
\bibliography{cas-refs}

\end{document}